\documentclass[prd,aps,showpacs,epsf,floats,onecolumn]{revtex4}%
\usepackage{amssymb}
\usepackage{amsfonts}
\usepackage{amsmath}
\usepackage{graphicx}%
\providecommand{\U}[1]{\protect\rule{.1in}{.1in}}
\begin{document}
\title{\textbf{Constructions of Optimal-Speed Quantum Evolutions: A Comparative
Study}}
\author{\textbf{Leonardo Rossetti}$^{1,2}$, \textbf{Carlo Cafaro}$^{2,3}$,
\textbf{Newshaw Bahreyni}$^{4}$}
\affiliation{$^{1}$University of Camerino, I-62032 Camerino, Italy}
\affiliation{$^{2}$University at Albany-SUNY, Albany, NY 12222, USA}
\affiliation{$^{3}$SUNY Polytechnic Institute, Utica, NY 13502, USA}
\affiliation{$^{4}$Pomona College, Claremont, CA 91711, USA}

\begin{abstract}
We present a comparative analysis of two different constructions of
optimal-speed quantum Hamiltonian evolutions on the Bloch sphere. In the first
approach (Mostafazadeh's approach), the evolution is specified by a traceless
stationary Hermitian Hamiltonian and occurs between two arbitrary qubit states
by maximizing the energy uncertainty. In the second approach (Bender's
approach), instead, the evolution is characterized by a stationary Hermitian
Hamiltonian which is not traceless and occurs between an initial qubit state
on the north pole and an arbitrary final qubit state. In this second approach,
the evolution occurs by minimizing the evolution time subject to the
constraint that the difference between the largest and the smallest
eigenvalues of the Hamiltonian is kept fixed. For both approaches\textbf{ }we
calculate explicitly the optimal Hamiltonian, the optimal unitary evolution
operator and, finally, the optimal magnetic field configuration. Furthermore,
we show in a clear way that Mostafazadeh's and Bender's approaches are
equivalent when we extend Mostafazadeh's approach to Hamiltonians with nonzero
trace and, at the same time, focus on an initial quantum state placed on the
north pole of the Bloch sphere. Finally, we demonstrate in both scenarios that
the optimal unitary evolution operator is a rotation about an axis that is
orthogonal to the unit Bloch vectors that correspond to the initial and final
qubit states.

\end{abstract}

\maketitle

\section{Introduction}

It is known that the geometry on the space of quantum states
\cite{karol,bohm03,dario04,avron09,avron20,anza21,anza22,anza22B,anza24,beggs20,majid19,beggs21,beggs22}%
, either pure \cite{wootters81} or mixed \cite{braunstein94}, can be used to
describe our limited capability in distinguishing one state from another in
terms of measurements. Interestingly, the initial approach to studying quantum
evolutions was taken in Refs. \cite{page87,abe92}. A geometric phase factor
emerges when a quantum state evolves around a closed path in the projective
Hilbert space of rays, and this phase is explicitly expressed through the
holonomies of natural geometrical structures on the projective space in Ref.
\cite{page87}. In Ref. \cite{abe92}, the study focuses on the Fubini-Study
metric induced on the quantum evolution submanifold of the projective Hilbert
space associated with the evolutions generated by a set of independent
operators. It is demonstrated that the off-diagonal components of the metric
describe the correlations between the Hermitian operators, giving a
geometrical interpretation to both uncertainty and correlation. Furthermore,
we also remark that aspects of time minimal trajectories for spin-$1/2$ and
higher spin particles in a magnetic field were previously investigated in
Refs. \cite{boscain06,boozer12} \ and Refs. \cite{fry08,kuz15},
respectively.\textbf{ }In general, the actual dynamical evolution of a quantum
system is not specified by the geometry on the space of quantum states
\cite{braunstein95}. In reality, not all Hamiltonian evolutions are shortest
time Hamiltonian evolutions. For this reason, actual quantum dynamical
trajectories traced out by points in projective Hilbert spaces differ from the
geodesic paths on the underlying quantum state space equipped with a suitable
metric (i.e., for pure states, the projective Hilbert space equipped with the
Fubini-Study metric). However, limiting our attention to pure states, there
are Hamiltonian operators generating optimal-speed quantum evolutions
specified by the shortest temporal duration
\cite{brody03,carlini06,brody06,brody07,bender07,bender09} together with the
maximal energy dispersion \cite{uhlmann92,ali09}. For such quantum motions,
Hamiltonian curves, viewed as quantum dynamical trajectories traced out by
quantum states evolving according to given physical evolutions, can be shown
to be geodesic paths on the underlying metricized manifolds of quantum states
\cite{cafaro23}. When analyzing unit geodesic efficiency
\cite{anandan90,cafaro20,hamilton23} quantum mechanical unitary evolutions
characterized by stationary Hamiltonians under which an initial unit state
vector $\left\vert A\right\rangle $ evolves into a final unit state vector
$\left\vert B\right\rangle $, two main alternative approaches emerge in the
literature \cite{ali09,bender07}. In the first approach by Mostafazadeh in
Ref. \cite{ali09}, one searches for an expression of the Hamiltonian by
maximizing the energy uncertainty $\Delta E$ of the quantum system. This first
approach is justified by the proportionality between the angular speed $v$ of
the minimal-time evolution of the quantum system and the energy uncertainty
$\Delta E$, $v\overset{\text{def}}{=}ds_{\mathrm{FS}}/dt\propto$ $\Delta E$,
with $s_{\mathrm{FS}}$ denoting the Fubini-Study distance between the two
points in the projective Hilbert space $\mathcal{P}\left(  \mathcal{H}\right)
$ that specify the chosen initial and final states $\left\vert A\right\rangle
$ and $\left\vert B\right\rangle $, respectively. In the second approach by
Bender and collaborators in Ref. \cite{bender07}, the goal is finding an
expression of the Hamiltonian by minimizing the evolution time $\Delta
t\overset{\text{def}}{=}(t_{B}-t_{A})$ needed for evolving from $\left\vert
A\right\rangle $ to $\left\vert B\right\rangle $ given that the difference
between the largest ($E_{+}$) and smallest ($E_{-}$) eigenvalues of the
Hamiltonian is maintained fixed (i.e., $E_{+}-E_{-}=\mathrm{fixed}$). Given
the fact that $\Delta E_{\max}=\left(  E_{+}-E_{-}\right)  /2$, upper bounding
the difference between the largest and the smallest eigenvalues of the
Hamiltonian is the same as upper bounding the energy uncertainty $\Delta E$.
Therefore,\textbf{ }one can reasonably expect that these two quantum
characterizations of geodesic Hamiltonian motion are essentially equivalent.
However, giving a closer look to them, one can recognize that these two
approaches put the emphasis on slightly distinct features. Interestingly, the
peculiar features of these two different ways to characterize geodesic quantum
evolutions were cleverly exploited in Ref. \cite{cafaro22} to help
characterizing the formal analogies between the geometry of quantum evolutions
with unit quantum geometric efficiency and the geometry of classical
polarization optics for light waves with degree of polarization\textbf{ }that
equals the degree of coherence between the electric vibrations in any two
mutually orthogonal directions of propagation of the wave \cite{wolf59,wolf07}%
. Although these distinctive features of Mostafazadeh's and Bender's
approaches were briefly pointed out in Ref. \cite{cafaro22}, a detailed
comparative analysis between the approaches in Ref. \cite{ali09} and Ref.
\cite{bender07} is missing in the literature. The main goal of this paper is
to fill this gap.

The rest of the paper is organized as follows. In Section II, we critically
revisit Mostafazadeh's approach as originally proposed in Ref. \cite{ali09}.
Specifically, we obtain the exact expressions of the optimal Hamiltonian, the
optimal evolution operator, and the optimal stationary magnetic field
configuration for a geodesic evolution on the Bloch sphere for qubits. In
Section III, we critically revisit Bender's approach as originally suggested
in Ref. \cite{bender07}. Particularly, in analogy to what accomplished for the
revisitation of Mostafazadeh's approach, we get the exact expressions of the
optimal Hamiltonian, the optimal evolution operator, and the optimal
stationary magnetic field configuration for geodesic motion on the Bloch
sphere for a two-level quantum system. In Section IV, we show that
Mostafazadeh's and Bender's approaches are equivalent when we extend
Mostafazadeh's approach to Hamiltonians with nonzero trace and, at the same
time, focus on an initial quantum state located on the north pole of the Bloch
sphere. Our concluding remarks appear in Section V. Finally, some technical
details are placed in Appendix A.

\section{The Mostafazadeh approach}

In this section, we present a critical revisitation of Mostafazadeh's approach
as originally presented in Ref. \cite{ali09}. In particular, we derive the
exact expressions of the optimal (Hermitian) Hamiltonian, the optimal
(unitary) evolution operator, and the optimal stationary magnetic field
configuration for geodesic motion on the Bloch sphere for a two-level quantum
system. For a review concerning the basic geometry on the Bloch sphere, we
refer to Refs. \cite{karol} and \cite{cafaro23}.

Before proceeding with our analysis, we would like to briefly clarify the link
among the concepts of state vectors, wave functions, and rays in quantum
mechanics. First of all, we describe in this paper the state of a two-level
quantum system by means of a state vector. Second of all, one can show that
the wave function $\Psi\left(  x\right)  $ describing the state of a general
quantum system can be replaced by a vector $\left\vert \Psi\right\rangle $
that belongs to a complete, normed, infinite-dimensional vector space (i.e., a
Hilbert space). In particular, such a vector $\left\vert \Psi\right\rangle $
encodes the same information as the original wave function $\Psi\left(
x\right)  $. Then, the latter can be understood as the $x$-th component of the
vector with respect to the basis $\left\{  \left\vert x\right\rangle \right\}
$ formed by the eigenvectors of the position operator $\hat{x}$, namely
$\Psi\left(  x\right)  \overset{\text{def}}{=}\left\langle x\left\vert
\Psi\right.  \right\rangle $. Lastly, physical states are represented by rays
of the Hilbert space and two state vectors $\left\vert \psi\left(  s\right)
\right\rangle $ and $\left\vert \psi^{\prime}\left(  s\right)  \right\rangle
\overset{\text{def}}{=}e^{i\alpha\left(  s\right)  }\left\vert \psi\left(
s\right)  \right\rangle $ parametrized by a parameter $s$ (in $%
\mathbb{R}
^{n}$, in general) define the same point on the manifolds of rays (i.e., the
Bloch sphere $S^{2}\cong%
\mathbb{C}
P^{1}$, with $%
\mathbb{C}
P^{1}$ denoting the projective Hilbert space that corresponds to the
two-dimensional complex Hilbert space $\mathcal{H}_{2}^{1}$ of single-qubit
quantum states). For further technical details on these aspects, we refer to
Refs. \cite{provost80,mukunda93}.

Returning to our main analysis, recall that the infinitesimal Fubini-Study
line element between two\textbf{ }neighbouring quantum states $\left\vert
\psi\left(  t\right)  \right\rangle $ and $\left\vert \psi\left(  t+dt\right)
\right\rangle $ is given by $ds_{\mathrm{FS}}^{2}=(1/4)ds^{2}=1-\left\vert
\left\langle \psi\left(  t\right)  \left\vert \psi\left(  t+dt\right)
\right.  \right\rangle \right\vert ^{2}=(\Delta E^{2}(t)/\hbar^{2}%
)dt^{2}=(v_{\mathrm{H}}^{2}(t)/4)dt^{2}$ \cite{anandan90}. Note that
$v_{\mathrm{H}}\left(  t\right)  $ denotes the speed of the quantum evolution
in projective Hilbert space, $s$ is the distance along the effective
(nongeodesic, in general) dynamical path that joins the initial ($\left\vert
A\right\rangle $) and final ($\left\vert B\right\rangle $) states, and
$s_{0}=2s_{\mathrm{FS}}=2\arccos(\left\vert \left\langle A\left\vert B\right.
\right\rangle \right\vert )$ is the (geodesic) distance along the shortest
geodesic joining the initial and final states. The Fubini-Study distance
$s_{\mathrm{FS}}$ is equal to one half of the geodesic distance $s_{0}$. The
factor $2$\textbf{ }between the geodesic distance\textbf{ }$s_{0}$\textbf{
}and the Fubini-Study one\textbf{ }$s_{\mathrm{FS}}$ depends on the fact that
the Fubini-Study distance can be interpreted as the angle whose\textbf{ }%
$\cos$ corresponds to the modulus of the scalar product between the
neighbouring states under consideration. But now we can easily see the need of
a factor\textbf{ }$2$\textbf{ }between the Fubini-Study distance and the usual
distance on the manifold the states belong to. Indeed if we consider qubit
states, the usual distance on the Bloch sphere between antipodal vectors is
$\pi$; at the same time the scalar product between states corresponding to
antipodal Bloch vectors is null, then the Fubini-Study distance between them
must be $\pi/2$. The quantum evolution is geodesic when $s_{0}/s=1$, with
$0\leq\eta_{\mathrm{GE}}\overset{\text{def}}{=}s_{0}/s\leq1$ being the
so-called geodesic efficiency \cite{anandan90}. For instance, for a geodesic
quantum evolution specified by a stationary Hamiltonian, $s_{0}%
=s=v_{\mathrm{H}}^{\max}\Delta t_{\min}=\left[  (2\cdot\Delta E_{\max}%
)/\hbar\right]  \Delta t_{\min}$. Roughly speaking, $s=v_{\mathrm{H}}^{\max
}\Delta t_{\min}$ means that if we assume evolving the state on the manifold
of quantum states, then if the speed of the evolution is bounded (i.e.,
$v_{\mathrm{H}}\leq v_{\mathrm{H}}^{\max}$), and we can always travel at the
maximum speed (i.e., at $v_{\mathrm{H}}^{\max}$), then we get from $\left\vert
A\right\rangle $ to $\left\vert B\right\rangle $ fastest (i.e., in minimal
time $\Delta t_{\min}$) by taking the shortest route (i.e., $s=s_{0}$). Before
moving to our first subsection, we remark that in Ref. \cite{ali09}, $\left(
ds^{2}\right)  _{\mathrm{Mostafazadeh}}=(1/4)\left(  ds^{2}\right)
_{\mathrm{Anandan-Aharonov}}$ (see Eqs. (2) and (7) in Refs. \cite{ali09} and
\cite{anandan90}, respectively). For this reason, keep in mind in what follows
that $2s_{\mathrm{Mostafazadeh}}=s_{\mathrm{Anandan-Aharonov}}$.

\subsection{The optimal Hamiltonian}

In what follows, we assume to consider the evolution of single qubit quantum
states. We want to find the time-independent Hamiltonian $\mathrm{H}$
maximizing the energy uncertainty $\Delta E_{\psi}$ or, alternatively,
minimizing the time interval $\tau$ needed to evolve from $\left\vert \psi
_{I}\right\rangle $ to $\left\vert \psi_{F}\right\rangle $. The equivalence
between the maximization of $\Delta E_{\psi}$ and the minimization of $\tau$
is justified by the existing relations between the time interval $\tau$, the
energy uncertainty $\Delta E_{\psi}$ and, finally, the distance $s$ traced by
the time evolution in the projective Hilbert space $\mathcal{P}(\mathcal{H})$.
The essential two relations are \cite{anandan90}%
\begin{equation}
s\overset{\text{def}}{=}\frac{1}{\hbar}\int_{0}^{\tau}\Delta E_{\psi
(t)}dt\text{, and }\Delta E_{\psi(t)}\overset{\text{def}}{=}\sqrt
{\frac{\langle\psi(t)|\mathrm{H}^{2}|\psi(t)\rangle}{\langle\psi
(t)|\psi(t)\rangle}-\frac{|\langle\psi(t)|\mathrm{H}|\psi(t)\rangle|^{2}%
}{\langle\psi(t)|\psi(t)\rangle^{2}}}\text{,}\label{eq1}%
\end{equation}
with $s=s_{\mathrm{Mostafazadeh}}=s_{\mathrm{Anandan-Aharonov}}/2$. When we
consider a time-independent Hamiltonian \textrm{H}, since the unitary
time-evolution operator $e^{-it\mathrm{H}/\hbar}$ commutes with $\mathrm{H}$
and $\mathrm{H}^{2}$, the energy uncertainty $\Delta E_{\psi(t)}$ does not
depend on time. Therefore, from the first relation in Eq. (\ref{eq1}), the
relation between $\tau$ and $s$ becomes%
\begin{equation}
\tau=\frac{\hbar s}{\Delta E_{\psi}}\text{.}\label{eq2}%
\end{equation}
Without loss of generality, we can focus on bidimensional and traceless
Hamiltonians. The bidimensionality assumption relies on the fact that the
shortest possible path (i.e., the geodesic) connecting $\left\vert \psi
_{I}\right\rangle $ and $\left\vert \psi_{F}\right\rangle $) lies entirely in
the projective Hilbert space. If in a linear space the distance is defined by
a norm, the metric is specified by the inner product \cite{crell09}.
Furthermore, the geodesics connecting two vectors $|\psi_{A}\rangle$ and
$|\psi_{B}\rangle$ are of the form $t\rightarrow\left\vert \psi
(t)\right\rangle \overset{\text{def}}{=}(1-t)|\psi_{A}\rangle+t|\psi
_{B}\rangle$ with $\left\vert \dot{\psi}(t)\right\rangle =|\psi_{B}%
\rangle-|\psi_{A}\rangle$; this explains why we can focus on bidimensional
space. The traceless condition, instead, will be explained in a better manner
in Appendix A. The traceless condition implies that the eigenvalues of
$\mathrm{H}$ must have opposite sign, that is to say $E_{2}=-E_{1}%
\overset{\text{def}}{=}E$. Let $\{|\psi_{1}\rangle$, $|\psi_{2}\rangle\}$ be
an orthonormal basis consisting of the eigenvectors of $\mathrm{H}$ with
$\mathrm{H}|\psi_{n}\rangle=E_{n}|\psi_{n}\rangle$. We can expand
$|\psi(0)\rangle=|\psi_{I}\rangle$ in this basis to find%
\begin{equation}
|\psi_{I}\rangle=c_{1}|\psi_{1}\rangle+c_{2}|\psi_{2}\rangle\text{,}%
\label{eq3}%
\end{equation}
with $c_{1}$, $c_{2}\in\mathbb{%
\mathbb{C}
}$. Moreover, exploiting the time independence of $\Delta E_{\psi}$, we can
compute it at $t=0$. Using Eqs. (\ref{eq1}) and (\ref{eq2}), we get
\begin{equation}
\Delta E_{\psi}=E\sqrt{1-{\left(  \frac{|c_{1}|^{2}-|c_{2}|^{2}}{|c_{1}%
|^{2}+|c_{2}|^{2}}\right)  }^{2}}\leq E\text{.}\label{eq4}%
\end{equation}
Therefore, from Eqs. (\ref{eq2}) and (\ref{eq4}), the travel time $\tau$
satisfies
\begin{equation}
\tau\geq\tau_{\text{\textrm{min}}}\overset{\text{def}}{=}\frac{\hbar s}%
{E}\text{,}\label{eq5}%
\end{equation}
where $s$ is the geodesic distance between the rays $\lambda_{|\psi_{I}%
\rangle}$ and $\lambda_{|\psi_{F}\rangle}$, respectively, corresponding to
$|\psi_{I}\rangle$ and $|\psi_{F}\rangle$ in the projective Hilbert space
$\mathcal{P}(\mathcal{H})$. Next, we construct the Hamiltonian with
eigenvalues $\pm E$ for which $\tau=\tau_{\text{\textrm{min}}}$. Since $s$ is
completely determined by $\lambda_{|\psi_{I}\rangle}$ and $\lambda_{|\psi
_{F}\rangle}$, the condition $\tau=\tau_{\text{\textrm{min}}}$ is fulfilled if
and only if $\Delta E_{\psi}=E$. In view of (\ref{eq4}), this is equivalent to
$|c_{1}|=|c_{2}|$. If we then expand $|\psi_{F}\rangle$ in the basis
$\{|\psi_{1}\rangle$, $|\psi_{2}\rangle\}$, we get
\begin{equation}
|\psi_{F}\rangle=d_{1}|\psi_{1}\rangle+d_{2}|\psi_{2}\rangle\text{,}%
\label{eq6}%
\end{equation}
with $d_{1}$, $d_{2}\in\mathbb{%
\mathbb{C}
}$. Then, computing $\Delta E_{\psi}$ at $t=\tau$, we obtain Eq. (\ref{eq4})
with $(c_{1}$, $c_{2})$ replaced by $(d_{1}$, $d_{2})$. As a result, in order
to maintain $\Delta E_{\psi}=E$, we must have $|d_{1}|=|d_{2}|$.
Interestingly, we can provide a first geometric interpretation of these
conditions $|c_{1}|=|c_{2}|$ and $|d_{1}|=|d_{2}|$. Specifically, a generic
traceless Hermitian bidimensional operator can be always written as
$\mathrm{A}\overset{\text{def}}{=}\mathbf{v}\cdot\mathbf{\sigma}$ with
$\mathbf{v}\in\mathbb{%
\mathbb{R}
}^{3}$. Then, it is possible to show that the corresponding eigenstates can be
represented in the Bloch sphere by the two antipodal unit vectors $\hat{v}$
and $-\hat{v}$. Note that any other normalized state $|\psi\rangle$ can be
expressed as linear combination of these two states since they constitute an
orthonormal basis. In analogy to what happens with the usual computational
basis $\{|0\rangle$, $|1\rangle\}$, we can write $|\psi\rangle=\cos
(\theta/2){|\hat{v}\rangle}+\sin(\theta/2){e^{i\phi}|-\hat{v}\rangle}$. The
quantity $\theta$ is the angle between the Bloch vector corresponding to
$|\psi\rangle$ and $\hat{v}$, while $\phi$ can be set arbitrarily as the
azimuthal angle. The time evolution operator represented by the unitary
operator $U=e^{-i\frac{t}{\hbar}\mathrm{A}}=e^{-i\frac{t}{\hbar}%
\mathbf{v}\cdot\mathbf{\sigma}}$ corresponds to a rotation around the axis
$\hat{v}$. Assuming normalized states, the conditions $|c_{1}|=|c_{2}%
|=1/\sqrt{2}$ and $|d_{1}|=|d_{2}|=1/\sqrt{2}$ will result in the condition
$\theta=\pi/2$, with theta the angle between\textbf{ }$|\psi_{I}\rangle
$\textbf{ }(or $|\psi_{F}\rangle$)\textbf{ }and the Bloch vector corresponding
to\textbf{ }$|\psi_{1}\rangle$. This, in turn, means that the (optimal)
rotation axis $\hat{v}$ must be chosen such that the final and initial states
belong to the azimuthal plane of the axis. This preliminary geometric
interpretation of Mostafazadeh's result exactly matches what we shall discuss
later in this paper. Assuming normalized states, we can set $c_{1}=1/\sqrt{2}%
$, $c_{2}=(1/\sqrt{2})e^{i\alpha_{I}}$, $d_{1}=1/\sqrt{2}$, and $d_{2}%
=(1/\sqrt{2})e^{i\alpha_{F}}$ with $\alpha_{I}$, $\alpha_{F}\in\mathbb{%
\mathbb{R}
}$. Substituting these relations in Eq. (\ref{eq3}) and in Eq. (\ref{eq6}), we
find
\begin{equation}
|\psi_{1}\rangle+e^{i\alpha_{I}}|\psi_{2}\rangle=\sqrt{2}|\psi_{I}%
\rangle\text{, and }\left\vert \psi_{1}\right\rangle +e^{i\alpha_{F}}|\psi
_{2}\rangle=\sqrt{2}|\psi_{F}\rangle\text{.}\label{eq7}%
\end{equation}
We can solve these relations in Eq. (\ref{eq7}) for $|\psi_{1}\rangle$ and
$|\psi_{2}\rangle$ in terms of $|\psi_{I}\rangle$ and $|\psi_{F}\rangle$.
Then, we can plug them into the spectral decomposition of the optimal
Hamiltonian
\begin{equation}
\mathrm{H}=E(-|\psi_{1}\rangle\langle\psi_{1}|+|\psi_{2}\rangle\langle\psi
_{2}|)\text{,}\label{eq8}%
\end{equation}
to find expression of $\mathrm{H}$ in terms of $|\psi_{I}\rangle$ and
$|\psi_{F}\rangle$. Explicitly, from the first condition in Eq. (\ref{eq7}),
we find
\begin{equation}
|\psi_{1}\rangle=\sqrt{2}|\psi_{I}\rangle-e^{i\alpha_{I}}|\psi_{2}%
\rangle\text{.}\label{eq9}%
\end{equation}
Then, inserting Eq. (\ref{eq9}) into second condition in Eq. (\ref{eq7}), we
obtain
\begin{equation}
|\psi_{2}\rangle=\sqrt{2}\frac{|\psi_{F}\rangle-|\psi_{I}\rangle}%
{e^{i\alpha_{F}}-e^{i\alpha_{I}}}.\label{eq10}%
\end{equation}
Plugging now Eq. (\ref{eq10}) into Eq. (\ref{eq9}), collecting $e^{i\alpha
_{I}}$ and defining $\theta\overset{\text{def}}{=}\alpha_{I}-\alpha_{F}$, we
get
\begin{equation}
|\psi_{1}\rangle=\sqrt{2}(\frac{|\psi_{F}\rangle-e^{-i\theta}|\psi_{I}\rangle
}{1-e^{-i\theta}})\text{, and }|\psi_{2}\rangle=\sqrt{2}\frac{|\psi_{I}%
\rangle-|\psi_{F}\rangle}{e^{i\alpha_{I}}(1-e^{-i\theta})}\text{.}%
\label{eq11a}%
\end{equation}
As a side remark, note that $\theta$ is the angular distance on the Bloch
sphere between the states $|\psi_{I}\rangle$ and $|\psi_{F}\rangle$. Indeed,
since both $\alpha_{I}$ and $\alpha_{F}$ represent azimuthal angles with
respect to the same axis (i.e., the axis $\hat{v}$) as previously mentioned,
they are angles belonging to the same plane. Then, their difference is the
angle between $|\psi_{I}\rangle$ and $|\psi_{F}\rangle$ since the
corresponding (Bloch) vectors entirely belong to the above mentioned azimuthal
plane. More explicitly,\textbf{ }$\theta=2\cos^{-1}\left[  \left\vert
\left\langle \psi_{I}\left\vert \psi_{F}\right.  \right\rangle \right\vert
\right]  =2\cos^{-1}\left[  \sqrt{(1+\hat{a}_{I}\cdot\hat{a}_{F})/2}\right]
$\textbf{, }where\textbf{ }$\hat{a}_{I}$\textbf{ }and\textbf{ }$\hat{a}_{F}%
$\textbf{ }are the Bloch vectors corresponding to\textbf{ }$|\psi_{I}\rangle
$\textbf{ }and\textbf{ }$|\psi_{F}\rangle$\textbf{, }respectively, such
that\textbf{ }$\hat{a}_{I}\cdot\hat{a}_{F}=\cos(\theta)$\textbf{.} To find
$\mathrm{H}$, we have to calculate the projectors $|\psi_{1}\rangle\langle
\psi_{1}|$ and $|\psi_{2}\rangle\langle\psi_{2}|$. Exploiting Eq.
(\ref{eq11a}), we get
\begin{equation}
|\psi_{1}\rangle\langle\psi_{1}|=2\frac{|\psi_{F}\rangle\langle\psi_{F}%
|+|\psi_{I}\rangle\langle\psi_{I}|-e^{-i\theta}|\psi_{I}\rangle\langle\psi
_{F}|-e^{i\theta}|\psi_{F}\rangle\langle\psi_{I}|}{(1-e^{-i\theta
})(1-e^{i\theta})}\text{,}\label{carlo1}%
\end{equation}
and,%
\begin{equation}
|\psi_{2}\rangle\langle\psi_{2}|=2\frac{|\psi_{F}\rangle\langle\psi_{F}%
|+|\psi_{I}\rangle\langle\psi_{I}|-|\psi_{I}\rangle\langle\psi_{F}|-|\psi
_{F}\rangle\langle\psi_{I}|}{(1-e^{-i\theta})(1-e^{i\theta})}\text{,}%
\label{carlo2}%
\end{equation}
respectively. Furthermore, using the trigonometric identities $1-e^{-i\theta
}=2ie^{-i\frac{\theta}{2}}\sin({\frac{\theta}{2})}$ and $1-e^{i\theta
}=-2ie^{i\frac{\theta}{2}}\sin({\frac{\theta}{2})}$, Eqs. (\ref{carlo1}) and
(\ref{carlo2}) yield%
\begin{align}
|\psi_{2}\rangle\langle\psi_{2}|-|\psi_{1}\rangle\langle\psi_{1}| &
=2\frac{|\psi_{I}\rangle\langle\psi_{F}|(e^{-i\theta}-1)+|\psi_{F}%
\rangle\langle\psi_{I}|(e^{i\theta}-1)}{(1-e^{-i\theta})(1-e^{i\theta}%
)}\nonumber\label{eq13}\\
&  =-2\left(  \frac{|\psi_{I}\rangle\langle\psi_{F}|}{1-e^{i\theta}}%
+\frac{|\psi_{F}\rangle\langle\psi_{I}|}{1-e^{-i\theta}}\right)  \nonumber\\
&  =\frac{i}{\sin({\frac{\theta}{2})}}(|\psi_{F}\rangle\langle\psi
_{I}|e^{i\frac{\theta}{2}}-|\psi_{I}\rangle\langle\psi_{F}|e^{-i\frac{\theta
}{2}})\text{.}%
\end{align}
Now, employing Eq. (\ref{eq7}), we get
\begin{equation}
\langle\psi_{I}|\psi_{F}\rangle=\frac{1}{2}(e^{-i\theta}+1)=e^{-i\frac{\theta
}{2}}\cos({\frac{\theta}{2})}\text{, and }\langle\psi_{F}|\psi_{I}%
\rangle=\frac{1}{2}(e^{i\theta}+1)=e^{i\frac{\theta}{2}}\cos({\frac{\theta}%
{2})}\text{,}%
\end{equation}
that is,
\begin{equation}
e^{i\frac{\theta}{2}}=\frac{\cos({\frac{\theta}{2})}}{\langle\psi_{I}|\psi
_{F}\rangle}\text{, and }e^{-i\frac{\theta}{2}}=\frac{\cos({\frac{\theta}{2}%
)}}{\langle\psi_{F}|\psi_{I}\rangle}\text{.}\label{carlo3}%
\end{equation}
Inserting Eq. (\ref{carlo3}) into Eq. (\ref{eq13}) and remembering the
spectral decomposition for $\mathrm{H}$ in Eq. (\ref{eq8}), we finally get
\begin{equation}
\mathrm{H}_{\mathrm{opt}}^{\mathrm{M}}=iE\cot({\frac{\theta}{2})}\left(
\frac{|\psi_{F}\rangle\langle\psi_{I}|}{\langle\psi_{I}|\psi_{F}\rangle}%
-\frac{|\psi_{I}\rangle\langle\psi_{F}|}{\langle\psi_{F}|\psi_{I}\rangle
}\right)  \text{,}\label{eq16}%
\end{equation}
which is exactly the expression of the optimal ($\mathrm{opt}$) Hamiltonian
found originally by Mostafazadeh ($\mathrm{M}$) in terms of the initial and
final states $|\psi_{I}\rangle$ and $|\psi_{F}\rangle$. In Fig. $1$, we depict
the set of orthonormal state vectors $\left\{  \left\vert \psi_{1}%
\right\rangle \text{, }\left\vert \psi_{2}\right\rangle \right\}  $ on the
Bloch sphere that specify the eigenstates of\textbf{ }$\mathrm{H}%
_{\mathrm{opt}}^{\mathrm{M}}$\textbf{ }in Eq. (\ref{eq16}) used to
evolve\textbf{ }$\left\vert \psi_{I}\right\rangle $\textbf{ }into\textbf{
}$\left\vert \psi_{F}\right\rangle $. Having found $\mathrm{H}_{\mathrm{opt}%
}^{\mathrm{M}}$, we are ready to find $U_{\mathrm{opt}}^{\mathrm{M}%
}(t)=e^{-\frac{i}{\hbar}\mathrm{H}_{\mathrm{opt}}^{\mathrm{M}}t}$ in the next
subsection.\begin{figure}[t]
\centering
\includegraphics[width=0.25\textwidth] {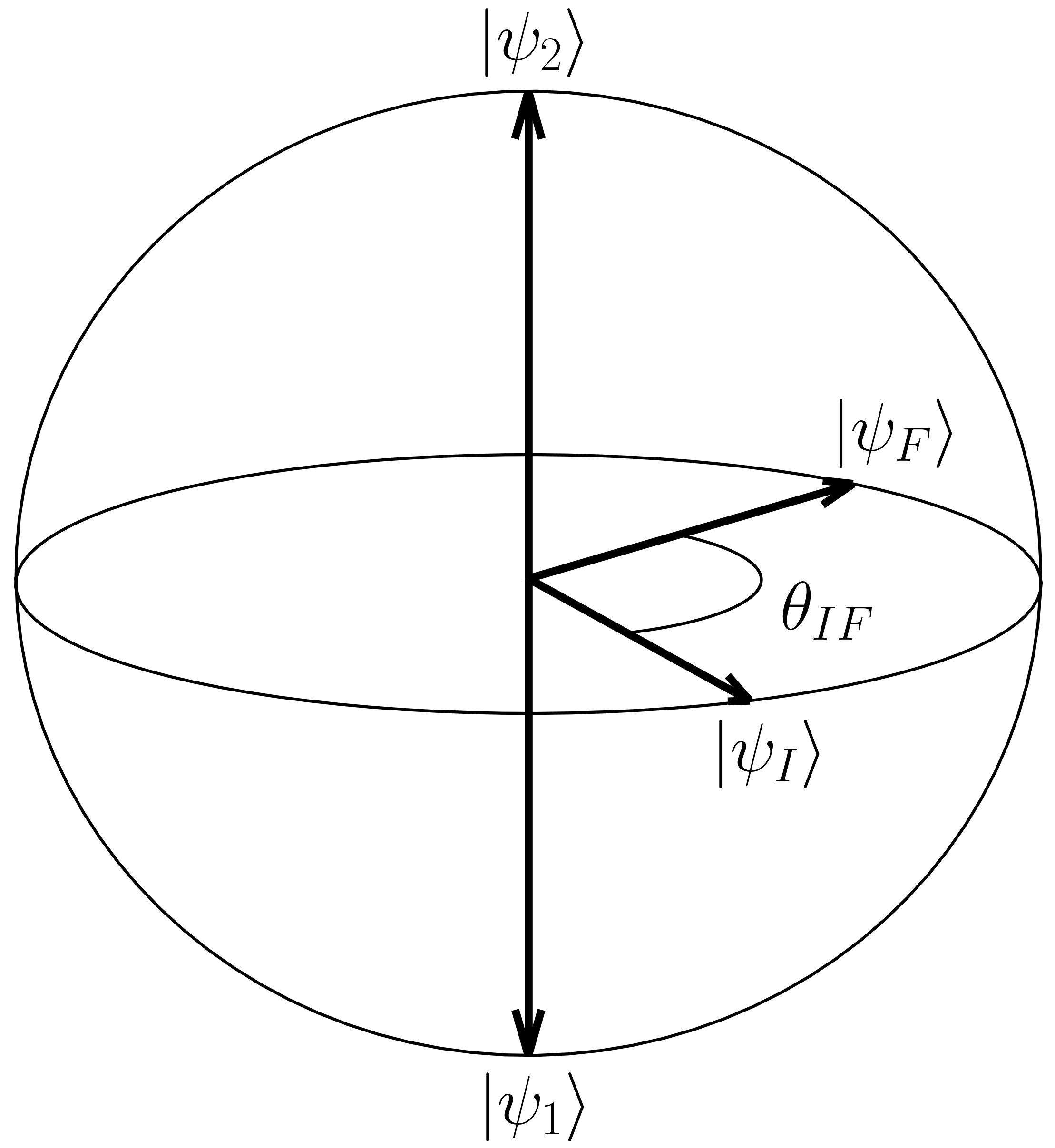}\caption{Schematic depiction of
the set of orthonormal state vectors $\left\{  \left\vert \psi_{1}%
\right\rangle \text{, }\left\vert \psi_{2}\right\rangle \right\}  $ on the
Bloch sphere that specify the eigenstates of the optimal Hamiltonian used to
evolve $\left\vert \psi_{I}\right\rangle $ into $\left\vert \psi
_{F}\right\rangle $ with $\left\vert \left\langle \psi_{I}\left\vert \psi
_{F}\right.  \right\rangle \right\vert =\cos(\theta_{IF}/2)$.}%
\end{figure}

\subsection{The optimal evolution operator}

From now on, we will refer to the initial and final states $|\psi_{I}\rangle$
and $|\psi_{F}\rangle$, respectively, with $|A\rangle$ and $|B\rangle$. From
the expression of the optimal Hamiltonian in Eq. (\ref{eq16}), we want to find
the corresponding unitary time evolution operator $U_{\mathrm{opt}%
}^{\mathrm{M}}(t)=e^{-\frac{i}{\hbar}\mathrm{H}_{\mathrm{opt}}^{\mathrm{M}}t}$
and verify in an explicit fashion that $|B\rangle=U_{\mathrm{opt}}%
^{\mathrm{M}}(\tau_{\text{\textrm{min}}})|A\rangle$. When we have the spectral
decomposition of an operator, a generic function of such operator will be
characterized by the same eigenvectors. However, the eigenvalues are given by
the application of this generic function on the eigenvalues of the original
operator. This means that if the spectral decomposition of $\mathrm{H}%
_{\mathrm{opt}}^{\mathrm{M}}$ is given by $\mathrm{H}_{\mathrm{opt}%
}^{\mathrm{M}}=E(-|\psi_{1}\rangle\langle\psi_{1}|+|\psi_{2}\rangle\langle
\psi_{2}|)$, then $U_{\mathrm{opt}}^{\mathrm{M}}(t)=e^{-i\frac{t}{\hbar
}\mathrm{H}_{\mathrm{opt}}^{\mathrm{M}}}$ is given by
\begin{equation}
U_{\mathrm{opt}}^{\mathrm{M}}(t)=e^{i\frac{t}{\hbar}E}|\psi_{1}\rangle
\langle\psi_{1}|+e^{-i\frac{t}{\hbar}E}|\psi_{2}\rangle\langle\psi
_{2}|\text{.} \label{eq17}%
\end{equation}
From the expressions for $|\psi_{1}\rangle\langle\psi_{1}|$ and $|\psi
_{2}\rangle\langle\psi_{2}|$ obtained in Eqs. (\ref{carlo1}) and
(\ref{carlo2}) and taking into account the fact that $1-e^{-i\theta
}=(1-e^{i\theta})^{\ast}$, we obtain%
\begin{align}
U_{\mathrm{opt}}^{\mathrm{M}}(t)  &  =\frac{2}{|1-e^{-i\theta}|^{2}}\left[
\begin{array}
[c]{c}%
e^{i\gamma}|B\rangle\langle B|+e^{i\gamma}|A\rangle\langle A|-e^{i(\gamma
-\theta)}|A\rangle\langle B|-e^{i(\gamma+\theta)}|B\rangle\langle A|+\\
+e^{-i\gamma}|B\rangle\langle B|+e^{-i\gamma}|A\rangle\langle A|-e^{-i\gamma
}|A\rangle\langle B|-e^{-i\gamma}|B\rangle\langle A|
\end{array}
\right] \nonumber\\
&  =\frac{2}{4\sin^{2}({\frac{\theta}{2})}}\left[  2\cos({\gamma)}%
|B\rangle\langle B|+2\cos({\gamma)}|A\rangle\langle A|-(e^{-i\gamma
}+e^{i(\gamma-\theta)})|A\rangle\langle B|-(e^{-i\gamma}+e^{i(\gamma+\theta
)})|B\rangle\langle A|\right] \nonumber\\
&  =\frac{1}{\sin^{2}({\frac{\theta}{2})}}\left[  \cos({\gamma)}%
|B\rangle\langle B|+\cos{\gamma}|A\rangle\langle A|-e^{-i\frac{\theta}{2}}%
\cos({\gamma-\frac{\theta}{2})}|A\rangle\langle B|-e^{i\frac{\theta}{2}}%
\cos({\gamma+\frac{\theta}{2})}|B\rangle\langle A|\right]  \text{,}%
\end{align}
that is,%
\begin{equation}
U_{\mathrm{opt}}^{\mathrm{M}}(t)=\frac{1}{\sin^{2}({\frac{\theta}{2})}}\left[
\cos({\gamma)}(|B\rangle\langle B|+|A\rangle\langle A|)-e^{-i\frac{\theta}{2}%
}\cos({\gamma-\frac{\theta}{2})}|A\rangle\langle B|-e^{i\frac{\theta}{2}}%
\cos({\gamma+\frac{\theta}{2})}|B\rangle\langle A|\right]  \text{.}
\label{eq19}%
\end{equation}
In the expression of $U_{\mathrm{opt}}^{\mathrm{M}}(t)$ in Eq. (\ref{eq19}),
$\theta\overset{\text{def}}{=}\alpha_{I}-\alpha_{F}$ and $\gamma
\overset{\text{def}}{=}(t/\hbar)E$. For completeness, we point out that to
obtain Eq. (\ref{eq19}), we used the fact that $|1-e^{-i\theta}|^{2}%
=|1-\cos({\theta)}+i\sin({\theta)}|^{2}=2-2\cos({\theta)}=4\sin^{2}%
(\theta/2{)}$, $e^{-i\gamma}+e^{i(\gamma-\theta)}=2e^{-i\frac{\theta}{2}}%
\cos({\gamma-\theta/2)}$, and $e^{-i\gamma}+e^{i(\gamma+\theta)}%
=2e^{i\frac{\theta}{2}}\cos({\gamma+\theta/2)}$. Having found $U_{\mathrm{opt}%
}^{\mathrm{M}}(t)$, we can verify if $|B\rangle=U_{\mathrm{opt}}^{\mathrm{M}%
}(\tau_{\text{\textrm{min}}})|A\rangle$. Note that $\tau_{\text{\textrm{min}}%
}$ equals $\tau_{\text{\textrm{min}}}=\hbar s_{\text{\textrm{FS}}}/\Delta
E_{\text{\textrm{max}}}$, with $s_{\text{\textrm{FS}}}$ being the Fubini-Study
distance in the projective Hilbert space $\mathcal{P}(\mathcal{H})$ between
the initial and final states. In our case, since the Fubini-Study distance on
the Bloch sphere corresponds to half the angular distance (i.e., the geodesic
distance) on the sphere, we can set $s_{\text{\textrm{FS}}}=\theta/2$. As a
consequence, for $t=\tau_{\text{\textrm{min}}}$, we get $\gamma=\theta/2$
since $\Delta E_{\text{\textrm{max}}}=E$. \noindent Plugging this value of
$\gamma$ into Eq. (\ref{eq19}), using Eq. (\ref{carlo3}), and applying the
resulting unitary time evolution operator$U_{\mathrm{opt}}^{\mathrm{M}}(t)$ to
$|A\rangle$, we get
\begin{align}
U_{\mathrm{opt}}^{\mathrm{M}}(\tau_{\text{\textrm{min}}})|A\rangle &
=\frac{1}{\sin^{2}({\frac{\theta}{2})}}\left[  \cos({\frac{\theta}{2}%
)}|B\rangle\langle B|A\rangle+\cos({\frac{\theta}{2})}|A\rangle-e^{-i\frac
{\theta}{2}}|A\rangle\langle B|A\rangle-e^{i\frac{\theta}{2}}\cos({\theta
)}|B\rangle\right] \nonumber\label{eq20}\\
&  =\frac{1}{\sin^{2}({\frac{\theta}{2})}}\left[  \cos^{2}({\frac{\theta}{2}%
)}e^{i\frac{\theta}{2}}-\cos^{2}({\frac{\theta}{2})}e^{i\frac{\theta}{2}}%
+\sin^{2}({\frac{\theta}{2})}e^{i\frac{\theta}{2}}\right]  |B\rangle
\nonumber\\
&  =e^{i\frac{\theta}{2}}|B\rangle\text{,}%
\end{align}
that is, $U_{\mathrm{opt}}^{\mathrm{M}}(\tau_{\text{\textrm{min}}}%
)|A\rangle=e^{i\frac{\theta}{2}}|B\rangle$. Since in quantum mechanics states
are equivalent up to an overall phase, $e^{i\frac{\theta}{2}}|B\rangle$ and
$|B\rangle$ are physically equivalent states. Therefore, we can safely
conclude that we explicitly verified that $U_{\mathrm{opt}}^{\mathrm{M}}%
(\tau_{\text{\textrm{min}}})|A\rangle=e^{i\frac{\theta}{2}}|B\rangle
\sim|B\rangle$. Having found $\mathrm{H}_{\mathrm{opt}}^{\mathrm{M}}$ and
$U_{\mathrm{opt}}^{\mathrm{M}}(t)=e^{-\frac{i}{\hbar}\mathrm{H}_{\mathrm{opt}%
}^{\mathrm{M}}t}$, in the next subsection we shall find the optimal magnetic
field configuration for the optimal Hamiltonian $\mathrm{H}_{\mathrm{opt}%
}^{\mathrm{M}}=\epsilon_{0}\mathbf{1}+\vec{\epsilon}\cdot\vec{\sigma}$ with
the magnetic field $\vec{B}$ proportional to the vector $\vec{\epsilon}$,
$\vec{B}\propto\vec{\epsilon}$.

\subsection{The optimal magnetic field}

In what follows, we want to express $\mathrm{H}_{\mathrm{opt}}^{\mathrm{M}}$
as $\mathrm{H}_{\mathrm{opt}}^{\mathrm{M}}=\epsilon_{0}\mathbf{1}%
+\vec{\epsilon}\cdot\vec{\sigma}$ with explicit expression of $\epsilon_{0}$
and $\vec{\epsilon}$. Since $\mathrm{H}_{\mathrm{opt}}^{\mathrm{M}}$ is
traceless, we expect to find $\epsilon_{0}=0$ and, for a suitable choice of
physical units, the vector $\vec{\epsilon}$ can be essentially viewed as the
(stationary) magnetic field vector in which the spin-$1/2$ particle (i.e., the
qubit) is immersed. For simplicity, we initially focus on the operator
$\frac{|B\rangle\langle A|}{\langle A|B\rangle}-\frac{|A\rangle\langle
B|}{\langle B|A\rangle}$ which is equal to $\mathrm{H}_{\mathrm{opt}%
}^{\mathrm{M}}$ in Eq. (\ref{eq16}) modulo a proportionality constant.
Expressing the states $|A\rangle$ and $|B\rangle$ as $|A\rangle
\overset{\text{def}}{=}a_{0}|0\rangle+a_{1}|1\rangle$ and $|B\rangle
\overset{\text{def}}{=}b_{0}|0\rangle+b_{1}|1\rangle$, we find that
\begin{align}
|B\rangle\langle A|  &  =b_{0}a_{0}^{\ast}|0\rangle\langle0|+b_{1}a_{0}^{\ast
}|1\rangle\langle0|+b_{0}a_{1}^{\ast}|0\rangle\langle1|+b_{1}a_{1}^{\ast
}|1\rangle\langle1|\text{,}\nonumber\\
|A\rangle\langle B|  &  =b_{0}^{\ast}a_{0}|0\rangle\langle0|+a_{1}b_{0}^{\ast
}|1\rangle\langle0|+a_{0}b_{1}^{\ast}|0\rangle\langle1|+a_{1}b_{1}^{\ast
}|1\rangle\langle1|\text{,}\nonumber\\
\langle A|B\rangle &  =a_{0}^{\ast}b_{0}+a_{1}^{\ast}b_{1}\text{,}\nonumber\\
\langle B|A\rangle &  =b_{0}^{\ast}a_{0}+b_{1}^{\ast}a_{1}\text{,}
\label{carlo4}%
\end{align}
where $a_{0}$, $a_{1}$, $b_{0}$, $b_{1}\in$ $%
\mathbb{C}
$. Using Eq. (\ref{carlo4}), $\frac{|B\rangle\langle A|}{\langle A|B\rangle
}-\frac{|A\rangle\langle B|}{\langle B|A\rangle}$ reduces to
\begin{align}
\frac{|B\rangle\langle A|}{\langle A|B\rangle}-\frac{|A\rangle\langle
B|}{\langle B|A\rangle}  &  =\left(  \frac{b_{0}a_{0}^{\ast}}{\langle
A|B\rangle}-\frac{a_{0}b_{0}^{\ast}}{\langle B|A\rangle}\right)
|0\rangle\langle0|+\left(  \frac{b_{1}a_{0}^{\ast}}{\langle A|B\rangle}%
-\frac{a_{1}b_{0}^{\ast}}{\langle B|A\rangle}\right)  |1\rangle\langle
0|+\nonumber\label{eq23}\\
&  +\left(  \frac{b_{0}a_{1}^{\ast}}{\langle A|B\rangle}-\frac{a_{0}%
b_{1}^{\ast}}{\langle B|A\rangle}\right)  |0\rangle\langle1|+\left(
\frac{b_{1}a_{1}^{\ast}}{\langle A|B\rangle}-\frac{a_{1}b_{1}^{\ast}}{\langle
B|A\rangle}\right)  |1\rangle\langle1|.
\end{align}
Expressing the coefficients of $|A\rangle$ and $|B\rangle$ as functions of the
usual (polar and azimuthal) angles for points on the Bloch sphere, we get
\begin{equation}
a_{0}=\cos({\frac{\theta_{A}}{2})}\text{, }a_{1}=\sin({\frac{\theta_{A}}{2}%
)}e^{i\varphi_{A}}\text{, }b_{0}=\cos({\frac{\theta_{B}}{2})}\text{, and
}b_{1}=\sin({\frac{\theta_{B}}{2})}e^{i\varphi_{B}}\text{.}%
\end{equation}
Exploiting the result in Eq. (\ref{carlo3}) to derive that $|\langle
A|B\rangle|^{2}=\cos^{2}({\frac{\theta}{2})}$, it follows that the
coefficients of the projectors $|0\rangle\langle0|$ and $|1\rangle\langle1|$
can be expressed as
\begin{equation}
\frac{b_{0}a_{0}^{\ast}}{\langle A|B\rangle}-\frac{a_{0}b_{0}^{\ast}}{\langle
B|A\rangle}=\frac{1}{\cos^{2}({\frac{\theta}{2})}}\cos({\frac{\theta_{A}}{2}%
)}\sin({\frac{\theta_{A}}{2})}\cos({\frac{\theta_{B}}{2})}\sin({\frac
{\theta_{B}}{2})}\cdot2i\cdot\sin{(\varphi_{A}-\varphi_{B})}\text{,}
\label{eq25}%
\end{equation}
and,%
\begin{equation}
\frac{b_{1}a_{1}^{\ast}}{\langle A|B\rangle}-\frac{a_{1}b_{1}^{\ast}}{\langle
B|A\rangle}=\frac{1}{\cos^{2}({\frac{\theta}{2})}}\cos({\frac{\theta_{A}}{2}%
)}\sin({\frac{\theta_{A}}{2})}\cos({\frac{\theta_{B}}{2})}\sin({\frac
{\theta_{B}}{2})}\cdot2i\cdot\sin{(\varphi_{B}-\varphi_{A})}\text{,}
\label{carlo5}%
\end{equation}
respectively. Moreover, the coefficients of the operators $|0\rangle\langle1|$
and $|1\rangle\langle0|$ can be expressed as
\begin{equation}
\frac{b_{0}a_{1}^{\ast}}{\langle A|B\rangle}-\frac{a_{0}b_{1}^{\ast}}{\langle
B|A\rangle}=\frac{1}{\cos^{2}({\frac{\theta}{2})}}\left[
\begin{array}
[c]{c}%
\cos^{2}({\frac{\theta_{B}}{2})}\cos({\frac{\theta_{A}}{2})}\sin({\frac
{\theta_{A}}{2})}e^{-i\varphi_{A}}+\cos({\frac{\theta_{B}}{2})}\sin^{2}%
({\frac{\theta_{A}}{2})}\sin({\frac{\theta_{B}}{2})}e^{-i\varphi_{B}}+\\
-\cos^{2}({\frac{\theta_{A}}{2})}\cos({\frac{\theta_{B}}{2})}\sin
({\frac{\theta_{B}}{2})}e^{-i\varphi_{B}}-\cos({\frac{\theta_{A}}{2})}\sin
^{2}({\frac{\theta_{B}}{2})}\sin({\frac{\theta_{A}}{2})}e^{-i\varphi_{A}}%
\end{array}
\right]  \text{,} \label{carlo8}%
\end{equation}
and,%
\begin{equation}
\frac{b_{1}a_{0}^{\ast}}{\langle A|B\rangle}-\frac{a_{1}b_{0}^{\ast}}{\langle
B|A\rangle}=\frac{1}{\cos^{2}({\frac{\theta}{2})}}\left[
\begin{array}
[c]{c}%
\cos^{2}({\frac{\theta_{A}}{2})}\sin({\frac{\theta_{B}}{2})}\cos({\frac
{\theta_{B}}{2})}e^{i\varphi_{B}}+\sin^{2}({\frac{\theta_{B}}{2})}\cos
({\frac{\theta_{A}}{2})}\sin({\frac{\theta_{A}}{2})}e^{i\varphi_{A}}+\\
-\cos({\frac{\theta_{A}}{2})}\sin({\frac{\theta_{A}}{2})}\cos^{2}%
({\frac{\theta_{B}}{2})}e^{i\varphi_{A}}-\cos({\frac{\theta_{B}}{2})}%
\sin({\frac{\theta_{B}}{2})}\sin^{2}({\frac{\theta_{A}}{2})}e^{i\varphi_{B}}%
\end{array}
\right]  \text{,} \label{carlo9}%
\end{equation}
respectively. Inserting Eqs. (\ref{eq25}), (\ref{carlo5}), (\ref{carlo8}), and
(\ref{carlo9}) into $\mathrm{H}_{\mathrm{opt}}^{\mathrm{M}}$ in Eq.
(\ref{eq16}), we get $\mathrm{H}_{\mathrm{opt}}^{\mathrm{M}}=\epsilon
_{0}\mathbf{1}+\vec{\epsilon}\cdot\vec{\sigma}$ as%
\begin{equation}
\mathrm{H}_{\mathrm{opt}}^{\mathrm{M}}=\epsilon_{0}\mathbf{1}+\vec{\epsilon
}\cdot\vec{\sigma}=\frac{E}{\sin({\theta)}}\left[
\begin{array}
[c]{c}%
\left[  \cos({\theta_{B})}\sin({\theta_{A})}\sin(\varphi{_{A})}-\cos
({\theta_{A})}\sin({\theta_{B})}\sin(\varphi{_{B})}\right]  \sigma_{x}+\\
+\left[  \cos({\theta_{B})}\sin({\theta_{A})}\cos(\varphi{_{A})}-\cos
({\theta_{A})}\sin({\theta_{B})}\cos(\varphi{_{B})}\right]  \sigma_{y}+\\
+\left[  \sin(\varphi{_{B}-\varphi_{A})}\sin({\theta_{A})}\sin({\theta_{B}%
)}\right]  \sigma_{z}%
\end{array}
\right]  \text{.} \label{carlo10}%
\end{equation}
From Eq. (\ref{carlo10}), we get $\epsilon_{0}=0$, while $\vec{\epsilon}$
becomes%
\begin{equation}
\vec{\epsilon}=\frac{E}{\sin{\theta}}\left(
\begin{array}
[c]{c}%
\overbrace{\cos({\theta_{B})}}^{z_{B}}\overbrace{\sin({\theta_{A})}%
\sin(\varphi{_{A})}}^{y_{A}}-\overbrace{\cos({\theta_{A})}}^{z_{A}%
}\overbrace{\sin({\theta_{B})}\sin(\varphi{_{B})}}^{y_{B}}\text{,}\\
\overbrace{\cos({\theta_{B})}}^{z_{B}}\overbrace{\sin({\theta_{A})}%
\cos(\varphi{_{A})}}^{x_{A}}-\overbrace{\cos({\theta_{A})}}^{z_{A}%
}\overbrace{\sin({\theta_{B})}\cos(\varphi{_{B})}}^{x_{B}}\text{,}\\
\overbrace{\sin(\varphi{_{B}-\varphi_{A})}\sin({\theta_{A})}\sin({\theta_{B}%
)}}^{x_{A}y_{B}-y_{A}x_{B}}%
\end{array}
\right)  \text{.} \label{carlo11b}%
\end{equation}
Setting $\hat{a}\overset{\text{def}}{=}(x_{A}$, $y_{A}$, $z_{A})=(\sin\left(
\theta_{A}\right)  \cos\left(  \varphi_{A}\right)  $, $\sin\left(  \theta
_{A}\right)  \sin\left(  \varphi_{A}\right)  $, $\cos(\theta_{A}))$ and
$\hat{b}\overset{\text{def}}{=}(x_{B}$, $y_{B}$, $z_{B})=(\sin\left(
\theta_{B}\right)  \cos\left(  \varphi_{B}\right)  $, $\sin\left(  \theta
_{B}\right)  \sin\left(  \varphi_{B}\right)  $, $\cos(\theta_{B}))$, we
note\textbf{ }that the vector $\vec{\epsilon}$ in Eq. (\ref{carlo11b}) is
proportional to the cross product between the unit vectors $\hat{a}$ and
$\hat{b}$ (i.e., the Bloch vectors corresponding to the initial and final
states $\left\vert A\right\rangle $ and $\left\vert B\right\rangle $,
respectively, with $\hat{a}\cdot$ $\hat{b}=\cos(\theta)$). Specifically, we
have $\vec{\epsilon}=\frac{E}{\sin{\theta}}(\hat{a}\times\hat{b})$. Therefore,
the optimal (Hermitian) Hamiltonian $\mathrm{H}_{\mathrm{opt}}^{\mathrm{M}}$
can be finally recast as
\begin{equation}
\mathrm{H}_{\mathrm{opt}}^{\mathrm{M}}=\frac{E}{\sin({\theta)}}(\hat{a}%
\times\hat{b})\cdot\vec{\sigma}\text{,} \label{eq30}%
\end{equation}
while its corresponding (unitary) time evolution operator becomes
\begin{equation}
U_{\mathrm{opt}}^{\mathrm{M}}(t)=\cos({\frac{Et}{\hbar})}\mathbf{1}%
-i\sin({\frac{Et}{\hbar})(}\frac{\hat{a}\times\hat{b}}{\sin({\theta)}}%
)\cdot\vec{\sigma}\text{.} \label{eq31}%
\end{equation}
Eq. (\ref{eq31}) implies that the time optimal evolution on the Bloch sphere
results in a rotation around the axis $\hat{a}\times\hat{b}/\left\Vert \hat
{a}\times\hat{b}\right\Vert =(\hat{a}\times\hat{b})/\sin({\theta)}$
perpendicular to the unit Bloch vectors $\hat{a}$ and $\hat{b}$ corresponding
to the initial and final states, respectively. In other words, the time
optimal evolutions is a rotation in a plane passing through such vectors and
the origin. Indeed, it is possible to show that the plane passing through two
given vectors and the origin is the one perpendicular to the vector resulting
from their cross product. Moreover, since a rotation always occurs in the
plane orthogonal to the rotation axis, this must be the one passing through
the origin. Finally, we refer to Appendix A for an explicit discussion of the
fact that the choice of focusing on traceless Hamiltonians does not affect the
generality of Mostafazadeh's approach to finding optimal-speed quantum
Hamiltonian evolutions. In Fig. $2$, we present a schematic depiction of the
action of $\mathrm{H}_{\mathrm{opt}}^{\mathrm{M}}$ in Eq. (\ref{eq30}) for
initial and final unit Bloch vectors $\hat{a}$ and $\hat{b}$, respectively, on
a Bloch sphere. \begin{figure}[t]
\centering
\includegraphics[width=0.35\textwidth] {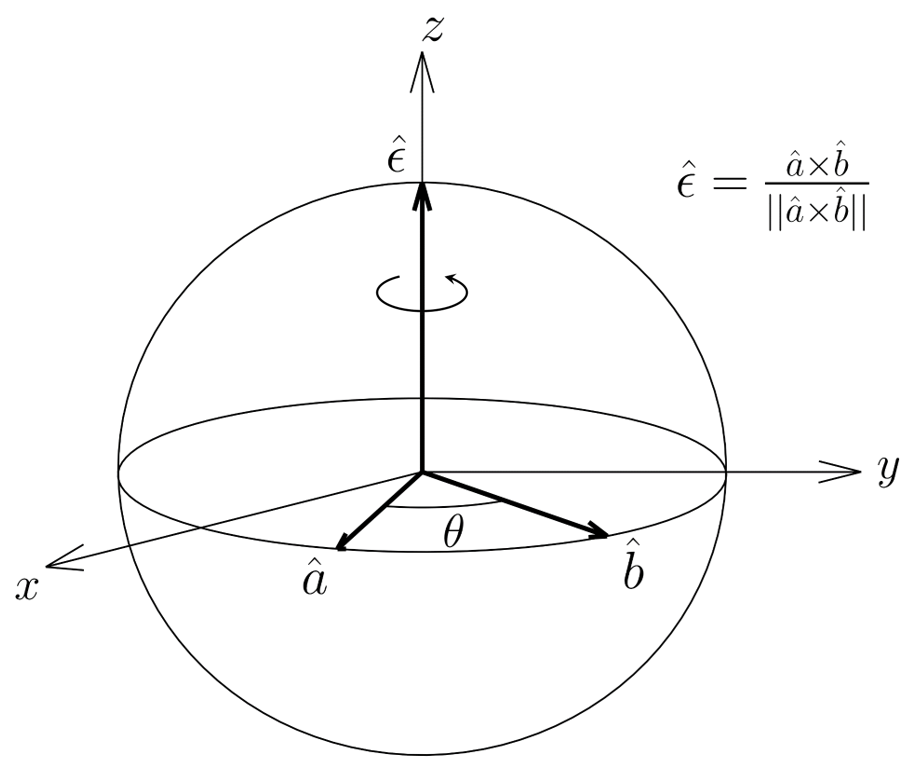}\caption{Schematic depiction of a
Bloch sphere where $\hat{a}$ and $\hat{b}$ are the initial and final unit
Bloch vectors with $\hat{a}\cdot\hat{b}=\cos(\theta)$. Furthermore,
$\hat{\epsilon}\protect\overset{\text{def}}{=}$ $(\hat{a}\times\hat{b}%
)/\sin(\theta)$ represents the unit vector that characterizes the axis of
rotation for the optimal-time Hamiltonian \textrm{H}$_{\mathrm{opt}%
}\protect\overset{\text{def}}{=}E\hat{\epsilon}\cdot\vec{\sigma}$ which serves
to evolve $\hat{a}$ into $\hat{b}$.}%
\end{figure}We are now ready to discuss Bender's approach to quantum geodesic
motion on a Bloch sphere with stationary (Hermitian) Hamiltonian evolutions.

\section{The Bender approach}

In this section, we present a critical revisitation of Bender's approach as
originally presented in Ref. \cite{bender07}. In particular, we derive the
exact expressions of the optimal (Hermitian) Hamiltonian, the optimal
(unitary) evolution operator, and the optimal stationary magnetic field
configuration for geodesic motion on the Bloch sphere for a two-level quantum system.

\subsection{The optimal Hamiltonian}

Consider initial and final quantum states given by $|A\rangle=\left(  1\text{,
}0\right)  ^{\mathrm{T}}$ and $|B\rangle=\left(  a\text{, }b\right)
^{\mathrm{T}}$, respectively, with $|a|^{2}+|b|^{2}=1$ and $a$, $b\in%
\mathbb{C}
$. We wish to find the optimal Hamiltonian $\mathrm{H}_{\text{\textrm{opt}}}$
that evolves the state $|A\rangle$ into the state $|B\rangle$ in the least
amount of time $\tau_{\text{\textrm{min}}}$ under the constraint that the
difference between the largest and the smallest eigenvalues of the Hamiltonian
is fixed. The most general $2\times2$ Hamiltonian can be expressed as
\begin{equation}
\mathrm{H}=%
\begin{pmatrix}
s & re^{-i\theta}\\
re^{i\theta} & u
\end{pmatrix}
\text{,}\label{eq64}%
\end{equation}
where the four parameters $r$, $s$, $u$, and $\theta$ are real. For clarity,
we stress that the parameter $s$ in Eq. (\ref{eq64}) used by Bender in Ref.
\cite{bender07} has energy units and is not the (adimensional) geodesic
distance parameter $s$ used by Mostafazadeh in Ref. \cite{ali09}. Moreover,
the parameter $\theta$ in Eq. (\ref{eq64}) does not denote the angular
distance on the Bloch sphere between the states $|\psi_{I}\rangle$ and
$|\psi_{F}\rangle$ as in Mostafazadeh's approach. We shall see later that, for
the optimal Hamiltonian originating from Eq. (\ref{eq64}), $\theta$ in Eq.
(\ref{eq64}) becomes\textbf{ }$\theta_{\mathrm{opt}}=\varphi_{B}+\pi
/2$\textbf{ }with\textbf{ }$\varphi_{B}$ being the azimuthal angle that
specifies the Bloch vector that corresponds to the final state\textbf{
}$|B\rangle$. The eigenvalue constraint $E_{+}-E_{-}=\omega$ for the
Hamiltonian in Eq. (\ref{eq64}) reduces to
\begin{equation}
\omega^{2}=(s-u)^{2}+4r^{2}\text{.}\label{eq65}%
\end{equation}
Indeed, if we consider the eigenvalue equation $\det\left(  \mathrm{H}%
-\lambda\mathbf{1}\right)  =0$ with $\mathrm{H}$ defined in Eq. (\ref{eq64}),
we get
\begin{equation}
\lambda_{1,2}\overset{\text{def}}{=}\frac{s+u\pm\sqrt{(s-u)^{2}+4r^{2}}}%
{2}\text{.}\label{eq66}%
\end{equation}
Hence $\lambda_{1}-\lambda_{2}=$ $E_{+}-E$ becomes%
\begin{equation}
E_{+}-E_{-}=\omega=\sqrt{(s-u)^{2}+4r^{2}}\text{.}\label{eq67}%
\end{equation}
At this point, we want to express the Hamiltonian $\mathrm{H}$ in Eq.
(\ref{eq64}) by means of the Pauli matrices in order to exploit the relation
$e^{-i\phi\hat{n}\cdot\vec{\sigma}}=\cos({\phi)}\mathbf{1}-i\sin({\phi)}%
\hat{n}\cdot\vec{\sigma}$ to connect the initial and final states $|A\rangle$
and $|B\rangle$. Recall that if $|B\rangle$ is obtained by time evolving
$|A\rangle$ under the action of the Hamiltonian $\mathrm{H}$ for a time $t$,
then it holds $|B\rangle=e^{-\frac{i}{\hbar}\mathrm{H}t}|A\rangle$. With the
help of some algebraic manipulations, we observe that
\begin{align}
\mathrm{H} &  =%
\begin{pmatrix}
s & r\cos({\theta)}-ir\sin({\theta)}\\
r\cos({\theta)}+ir\sin({\theta)} & u
\end{pmatrix}
\nonumber\label{eq68}\\
&  =%
\begin{pmatrix}
\frac{1}{2}(s+u)+\frac{1}{2}(s-u) & r\cos({\theta)}-ir\sin({\theta)}\\
r\cos({\theta)}+ir\sin({\theta)} & \frac{1}{2}(s+u)-\frac{1}{2}(s-u)
\end{pmatrix}
\nonumber\\
&  =\frac{1}{2}(s+u)\mathbf{1}+\overbrace{\frac{1}{2}(s-u)}^{a_{z}}\sigma
_{z}+\overbrace{r\cos({\theta)}}^{a_{x}}\sigma_{x}+\overbrace{r\sin({\theta)}%
}^{a_{y}}\sigma_{y}\nonumber\\
&  =\frac{1}{2}(s+u)\mathbf{1}+\frac{\omega}{2}\left[  \frac{2}{\omega}%
\cdot\frac{1}{2}(s-u)\sigma_{z}+\frac{2}{\omega}r\cos({\theta)}\sigma
_{x}+\frac{2}{\omega}r\sin({\theta)}\sigma_{y}\right]  \nonumber\\
&  =\frac{1}{2}(s+u)\mathbf{1}+\frac{\omega}{2}\hat{n}\cdot\vec{\sigma
}\text{,}%
\end{align}
that is,%
\begin{equation}
\mathrm{H}=\frac{1}{2}(s+u)\mathbf{1}+\frac{\omega}{2}\hat{n}\cdot\vec{\sigma
}\text{,}\label{carlo16}%
\end{equation}
with $\hat{n}\overset{\text{def}}{=}\frac{2}{\omega}(r\cos({\theta)}$,
$r\sin({\theta)}$, $\frac{s-u}{2})$. Note that to derive Eq. (\ref{carlo16}),
we made use of the relation $(a_{x}^{2}+a_{y}^{2}+a_{z}^{2})=(1/4)(s-u)^{2}%
+r^{2}=(1/4)\left[  (s-u)^{2}+4r^{2}\right]  =\omega^{2}/4$. We can now
calculate the unitary time evolution operator $U(t)=e^{-\frac{i}{\hbar
}\mathrm{H}t}$ determined by the Hamiltonian in Eq. (\ref{carlo16}). We have,
\begin{align}
e^{-\frac{i}{\hbar}\mathrm{H}t} &  =e^{-\frac{i}{\hbar}\left[  \frac{1}%
{2}(s+u)\mathbf{1}+\frac{\omega}{2}\hat{n}\cdot\vec{\sigma}\right]
t}\nonumber\label{eq69}\\
&  =e^{-\frac{i}{2\hbar}(s+u)\mathbf{1}t}e^{-i\frac{\omega}{2\hbar}\hat
{n}\cdot\vec{\sigma}t}\nonumber\\
&  =e^{-\frac{i}{2\hbar}(s+u)\mathbf{1}t}\left[  \cos({\frac{\omega}{2\hbar
}t)}-i\sin({\frac{\omega}{2\hbar}t)}\hat{n}\cdot\vec{\sigma}\right]
\nonumber\\
&  =e^{-\frac{i}{2\hbar}(s+u)t}%
\begin{pmatrix}
\cos({\frac{\omega}{2\hbar})t}-i\sin({\frac{\omega}{2\hbar}t)}\frac{2}{\omega
}\cdot\frac{(s-u)}{2} & \sin({\frac{\omega}{2\hbar}t)}\frac{2}{\omega}(\left[
-ir\cos({\theta)}-r\sin({\theta)}\right]  \\
\sin({\frac{\omega}{2\hbar}t)}\frac{2}{\omega}(\left[  -ir\cos({\theta)}%
+r\sin({\theta)}\right]   & \cos({\frac{\omega}{2\hbar}t)}+i\sin({\frac
{\omega}{2\hbar}t)}\frac{2}{\omega}\cdot\frac{(s-u)}{2}%
\end{pmatrix}
\text{,}%
\end{align}
that is,%
\begin{equation}
U(t)=e^{-\frac{i}{2\hbar}(s+u)t}%
\begin{pmatrix}
\cos({\frac{\omega}{2\hbar})t}-i\sin({\frac{\omega}{2\hbar}t)}\frac{2}{\omega
}\cdot\frac{(s-u)}{2} & \sin({\frac{\omega}{2\hbar}t)}\frac{2}{\omega}(\left[
-ir\cos({\theta)}-r\sin({\theta)}\right]  \\
\sin({\frac{\omega}{2\hbar}t)}\frac{2}{\omega}(\left[  -ir\cos({\theta)}%
+r\sin({\theta)}\right]   & \cos({\frac{\omega}{2\hbar}t)}+i\sin({\frac
{\omega}{2\hbar}t)}\frac{2}{\omega}\cdot\frac{(s-u)}{2}%
\end{pmatrix}
\text{.}\label{carlo17}%
\end{equation}
We can now apply this time evolution matrix in Eq. (\ref{carlo17}) to the
column representation of the initial state $|A\rangle$ and set it equal to the
column representation of the final state $|B\rangle$,
\begin{equation}
e^{-\frac{i}{\hbar}\mathrm{H}t}%
\begin{pmatrix}
1\\
0
\end{pmatrix}
=e^{-\frac{i}{2\hbar}(s+u)t}%
\begin{pmatrix}
\cos({\frac{\omega}{2\hbar}t)}-i\sin({\frac{\omega}{2\hbar}t)}\frac
{(s-u)}{\omega}\\
-i\sin({\frac{\omega}{2\hbar}t)}\frac{2r}{\omega}\left[  \cos({\theta)}%
+i\sin({\theta)}\right]
\end{pmatrix}
=%
\begin{pmatrix}
a\\
b
\end{pmatrix}
\text{.}\label{eq70}%
\end{equation}
From Eq. (\ref{eq70}), we obtain $|b|=(2r/\omega)\sin\left[  \left(
\omega/2\hbar\right)  {t}\right]  ${, that is}%
\begin{equation}
t=\frac{2\hbar}{\omega}\arcsin({\frac{\omega|b|}{2r})}\text{.}\label{eq71}%
\end{equation}
At this point, we need to minimize this expression for $t$ in Eq. (\ref{eq71})
to find the minimal time. Since $\omega$ is fixed and $\arcsin{(1/x)}$ is a
monotonic decreasing function of $x$, in order to get the smallest value for
$t$, we have to take the greatest possible value for $r$ in Eq. (\ref{eq71}).
Taking into account Eq. (\ref{eq65}), this value is given by $r=\omega/2$,
obtained when $s=u$. We observe that the condition $s=u$ corresponds to having
the $z$-component of the rotation axis in Eq. (\ref{eq68}) null, which means
that the rotation axis must belong to the $xy$-plane. This is in perfect
agreement with the fact that the Bloch vector corresponding to the initial
state is the $z$-versor. Indeed, the cross product between the $z$-versor and
any other versor will be a vector belonging to the $xy$-plane. Hence, we
obtained that in analogy to what we obtained in Eq. (\ref{eq30}), the optimal
rotation axis must be orthogonal to the cross product between the initial and
final unit Bloch vectors. Finally, the optimal time reduces to
\begin{equation}
\tau_{\min}=\frac{2\hbar}{\omega}\arcsin{|b|}\text{.}\label{eq72}%
\end{equation}
At the same time it results from Eq. (\ref{eq72}) that
\begin{equation}
|b|=\sin{\left(  \frac{\omega}{2\hbar}\tau_{\min}\right)  }\implies
\cos{\left(  \frac{\omega}{2\hbar}\tau_{\min}\right)  }=\sqrt{1-|b|^{2}%
}\text{.}\label{eq73}%
\end{equation}
Note that for $a=0$ and $b=1$ (or, in general, $b=e^{i\phi}$ with $\left\vert
b\right\vert =1$), one has the initial and final states that are mutually
orthogonal. In this case, $\tau_{\min}=\frac{2\pi\hbar}{\omega}$, is
called\emph{ passage time}. We can now calculate the explicit expression for
the time optimal Hamiltonian by inserting Eqs. (\ref{eq72}) and (\ref{eq73})
into Eq. (\ref{eq70}). We have,
\begin{equation}%
\begin{pmatrix}
a\\
b
\end{pmatrix}
=e^{-i\frac{s}{\hbar}\tau_{\min}}%
\begin{pmatrix}
\cos({\frac{\omega}{2\hbar}\tau_{\min})}\\
-i\sin({\frac{\omega}{2\hbar}\tau_{\min})}e^{i\theta}%
\end{pmatrix}
\implies%
\begin{pmatrix}
a\\
b
\end{pmatrix}
=e^{-i\frac{s}{\hbar}\tau_{\min}}%
\begin{pmatrix}
\sqrt{1-|b|^{2}}\\
-i|b|e^{i\theta}%
\end{pmatrix}
\text{.}\label{eq74}%
\end{equation}
In conclusion, recalling that $-i=e^{-i\frac{3}{2}\pi}$, we get from Eq.
(\ref{eq74}) that
\begin{equation}
a=e^{-i\frac{s}{\hbar}\tau_{\min}}\sqrt{1-|b|^{2}}\text{, and }b=|b|e^{i(\frac
{3}{2}\pi+\theta-\frac{s}{\hbar}\tau_{\min})}\text{.}%
\end{equation}
Since any complex number $c\in%
\mathbb{C}
$ can be written as $|c|e^{i\text{\textrm{arg}}(c)}$ we can express the
parameters of the Hamiltonian $s$ and $\theta$ in terms of \textrm{arg}$(a)$
and \textrm{arg}$(b)$. Using Eqs. (\ref{eq73}) and (\ref{eq74}), we have
\begin{equation}
s=-\frac{\mathrm{arg}(a)\hbar}{\tau_{\min}}=-\frac{\omega}{2}\frac
{\mathrm{arg}(a)}{\arcsin{|b|}}\text{, and }\theta=\mathrm{arg}(b)+\frac
{s}{\hbar}\tau_{\min}-\frac{3}{2}\pi=\mathrm{arg}(b)-\mathrm{arg}(a)-\frac
{3}{2}\pi\text{.}\label{carlo18}%
\end{equation}
In the end, keeping in mind that $s=u$ and $r=\omega/2$, use of Eqs.
(\ref{eq64}) and (\ref{carlo18}) yield the final expression for the optimal
Hamiltonian as originally obtained by Bender and collaborators,
\begin{equation}
\mathrm{H}_{\text{\textrm{opt}}}^{\mathrm{B}}=%
\begin{pmatrix}
-\frac{\omega}{2}\frac{\mathrm{arg}(a)}{\arcsin{|b|}} & \frac{\omega}%
{2}e^{-i\left[  \mathrm{arg}(b)-\mathrm{arg}(a)-\frac{3}{2}\pi\right]  }\\
\frac{\omega}{2}e^{i\left[  \mathrm{arg}(b)-\mathrm{arg}(a)-\frac{3}{2}%
\pi\right]  } & -\frac{\omega}{2}\frac{\mathrm{arg}(a)}{\arcsin{|b|}}%
\end{pmatrix}
\text{.}\label{eq77}%
\end{equation}
From Eq. (\ref{eq77}), we note that we get a traceless Hamiltonian
when\textrm{ arg}$(a)=0$. In this particular case, the traceless Hamiltonian
we obtain from Eq. (\ref{eq77}) does not have diagonal components. This is due
to the fact that the contributions to the diagonal components of a
bidimensional Hamiltonian result from the identity $\mathbf{1}$ and from
$\sigma_{z}$. However, the traceless condition is equivalent to not having
contributions from the identity as shown in Appendix A. Furthermore, as
previously explained, since the optimal rotation axis has no $z$-component, we
get no contribution to $\mathrm{H}_{\text{\textrm{opt}}}^{\mathrm{B}}$ from
$\sigma_{z}$. As a side remark, referring to the following section, we can
notice how imposing a constraint on the difference of the eigenvalues of the
Hamiltonian is perfectly equivalent to imposing a condition on the standard
deviation of the Hamiltonian. The reason behind this remark will appear more
transparent in the next section, when we explicitly show the correspondence
between the Mostafazadeh and the Bender approaches to optimal-speed quantum
Hamiltonian evolutions. Having found $\mathrm{H}_{\mathrm{opt}}^{\mathrm{B}}$,
we are now ready to obtain $U_{\mathrm{opt}}^{\mathrm{B}}(t)=e^{-\frac
{i}{\hbar}\mathrm{H}_{\mathrm{opt}}^{\mathrm{B}}t}$ together with the optimal
magnetic field configuration for the optimal Hamiltonian $\mathrm{H}%
_{\mathrm{opt}}^{\mathrm{B}}=\epsilon_{0}\mathbf{1}+\vec{\epsilon}\cdot
\vec{\sigma}$.

\subsection{The optimal evolution operator and magnetic field}

We wish to express $\mathrm{H}_{\text{\textrm{opt}}}^{\mathrm{B}}$ in Eq.
(\ref{eq77}) as $\mathrm{H}_{\text{\textrm{opt}}}^{\mathrm{B}}=\epsilon
_{0}\mathbf{1}+\vec{\epsilon}\cdot\vec{\sigma}$ with explicit expressions of
$\epsilon_{0}$ and $\vec{\epsilon}$. Then, given these expressions, we wish to
construct the unitary time evolution operator $U_{\mathrm{opt}}^{\mathrm{B}%
}(t)\overset{\text{def}}{=}e^{-\frac{i}{\hbar}\mathrm{H}_{\text{\textrm{opt}}%
}^{\mathrm{B}}t}$ given by
\begin{equation}
U_{\mathrm{opt}}^{\mathrm{B}}(t)=e^{-\frac{i}{\hbar}\left(  \epsilon
_{0}\mathbf{1}+\vec{\epsilon}\cdot\vec{\sigma}\right)  t}=e^{-\frac{i}{\hbar
}\epsilon_{0}t}\left[  \cos({\frac{\epsilon}{\hbar}t)}\mathbf{1}-i\sin
({\frac{\epsilon}{\hbar}t)}\hat{\epsilon}\cdot\vec{\sigma}\right]  \text{,}
\label{eq78}%
\end{equation}
with $\epsilon\overset{\text{def}}{=}\sqrt{\vec{\epsilon}\cdot\vec{\epsilon}}%
$. Finally, we wish to check that $|B\rangle=U_{\mathrm{opt}}^{\mathrm{B}%
}(\tau_{\mathrm{\min}})|A\rangle$ with $|A\rangle=\left(  1\text{, }0\right)
^{\mathrm{T}}$ and $|B\rangle=\left(  a\text{, }b\right)  ^{\mathrm{T}%
}=(\left\vert a\right\vert e^{i\mathrm{\arg}(a)}$, $\left\vert b\right\vert
e^{i\mathrm{\arg}(b)})^{\mathrm{T}}=(\cos(\theta_{B}/2)$, $\sin(\theta
_{B}/2)e^{i\varphi_{B}})^{\mathrm{T}}$. Let us start by manipulating the
expression for the optimal Hamiltonian $\mathrm{H}_{\text{\textrm{opt}}%
}^{\mathrm{B}}$ found in Eq. (\ref{eq77}). We note that,
\begin{align}
\mathrm{H}_{\text{\textrm{opt}}}^{\mathrm{B}}  &  =%
\begin{pmatrix}
-\frac{\omega}{2}\frac{\mathrm{arg}(a)}{\arcsin{|b|}} & \frac{\omega}%
{2}e^{-i\left[  \mathrm{arg}(b)-\mathrm{arg}(a)-\frac{3}{2}\pi\right]  }\\
\frac{\omega}{2}e^{i\left[  \mathrm{arg}(b)-\mathrm{arg}(a)-\frac{3}{2}%
\pi\right]  } & -\frac{\omega}{2}\frac{\mathrm{arg}(a)}{\arcsin{|b|}}%
\end{pmatrix}
\nonumber\\
&  =-\frac{\omega}{2}\frac{\text{\textrm{arg}}(a)}{\arcsin{|b|}}%
\mathbf{1}+\frac{\omega}{2}\cos({\theta)}\sigma_{x}+\frac{\omega}{2}%
\sin({\theta)}\sigma_{y}\nonumber\\
&  =\epsilon_{0}\mathbf{1}+\vec{\epsilon}\cdot\vec{\sigma}\text{,}
\label{carlo25}%
\end{align}
where $\theta=$\textrm{arg}$(b)-$\textrm{arg}$(a)-(3/2)\pi$,
\begin{equation}
\epsilon_{0}\overset{\text{def}}{=}-\frac{\omega}{2}\frac{\text{\textrm{arg}%
}(a)}{\arcsin{|b|}}\text{, and }\vec{\epsilon}\overset{\text{def}}{=}\left(
\frac{\omega}{2}\cos({\theta)}\text{, }\frac{\omega}{2}\sin({\theta)}\text{,
}0\right)  \text{.} \label{carlo19}%
\end{equation}
Hence, noticing from Eq. (\ref{carlo19}) that the modulus of $\vec{\epsilon}$
corresponds to $\epsilon=\omega/2$, we get the following expression for the
corresponding time evolution operator $U_{\mathrm{opt}}^{\mathrm{B}}(t)$ in
Eq. (\ref{eq78}),
\begin{equation}
U_{\mathrm{opt}}^{\mathrm{B}}(t)=e^{i\frac{\omega}{2\hbar}\frac
{\text{\textrm{arg}}(a)}{\arcsin{|b|}}t}\left\{  \cos({\frac{\omega}{2\hbar
}t)}\mathbf{1}-i\sin({\frac{\omega}{2\hbar}t)}\left[  \cos({\theta)}\sigma
_{x}+\sin({\theta)}\sigma_{y}\right]  \right\}  \text{.} \label{carlo20}%
\end{equation}
Inserting the value of $\tau_{\mathrm{min}}$ found in Eq. (\ref{eq72}) into
$U_{\mathrm{opt}}^{\mathrm{B}}(t)$ in Eq. (\ref{carlo20}), we obtain
\begin{align}
U_{\mathrm{opt}}^{\mathrm{B}}(\tau_{\mathrm{min}})  &  =e^{i\text{\textrm{arg}%
}(a)}\left\{  \cos({\arcsin{|b|)}}\mathbf{1}-i|b|\left[  \cos({\theta)}%
\sigma_{x}+\sin({\theta)}\sigma_{y}\right]  \right\} \nonumber\label{eq82}\\
&  =e^{i\text{\textrm{arg}}(a)}\left\{  \sqrt{1-\sin^{2}{(\arcsin{|b|})}%
}\mathbf{1}-i|b|\left[  \cos({\theta)}\sigma_{x}+\sin({\theta)}\sigma
_{y}\right]  \right\} \nonumber\\
&  =e^{i\text{\textrm{arg}}(a)}\left\{  \sqrt{1-|b|^{2}}\mathbf{1}-i|b|\left[
\cos({\theta)}\sigma_{x}+\sin({\theta)}\sigma_{y}\right]  \right\}  \text{,}%
\end{align}
that is,%
\begin{equation}
U_{\mathrm{opt}}^{\mathrm{B}}(\tau_{\mathrm{min}})=e^{i\text{\textrm{arg}}%
(a)}\left\{  \sqrt{1-|b|^{2}}\mathbf{1}-i|b|\left[  \cos({\theta)}\sigma
_{x}+\sin({\theta)}\sigma_{y}\right]  \right\}  \text{.} \label{carlo21}%
\end{equation}
We remark that given $\mathrm{H}_{\text{\textrm{opt}}}^{\mathrm{B}}%
=\epsilon_{0}\mathbf{1}+\vec{\epsilon}\cdot\vec{\sigma}$ in Eq. (\ref{carlo25}%
), we note that $\hat{\epsilon}=\vec{\epsilon}/\sqrt{\vec{\epsilon}\cdot
\vec{\epsilon}}=(\hat{a}\times\hat{b})/\sqrt{(\hat{a}\times\hat{b})\cdot
(\hat{a}\times\hat{b})}$. Indeed, in our case we have $\hat{a}%
\overset{\text{def}}{=}(0$, $0$, $1)$, $\hat{b}\overset{\text{def}}{=}%
(\sin\left(  \theta_{B}\right)  \cos\left(  \varphi_{B}\right)  $,
$\sin\left(  \theta_{B}\right)  \sin\left(  \varphi_{B}\right)  $,
$\cos(\theta_{B}))$, and $\hat{\epsilon}=\left(  -\sin\left(  \varphi
_{B}\right)  \text{, }\cos(\varphi_{B})\text{, }0\right)  $. However, given
that $\hat{\epsilon}_{\mathrm{Bender}}=\left(  \cos(\theta)\text{, }%
\sin(\theta)\text{, }0\right)  $, use of Eq. (\ref{carlo18}) together with the
fact that $-i=e^{-i\frac{\pi}{2}}=e^{i\frac{3}{2}\pi}$, we obtain that
$\theta$ equals $\theta_{\mathrm{opt}}\overset{\text{def}}{=}\varphi_{B}%
+\pi/2$. As a consequence, we get that $\hat{\epsilon}_{\mathrm{Bender}}%
=\hat{\epsilon}$. This geometrically meaningful result was not pointed out in
Ref. \cite{bender07}. Moreover, given $U_{\mathrm{opt}}^{\mathrm{B}}%
(\tau_{\mathrm{min}})$ in Eq. (\ref{carlo21}), we can finally verify in an
explicit manner that $|B\rangle=U_{\mathrm{opt}}^{\mathrm{B}}(\tau
_{\mathrm{\min}})|A\rangle$ with $|A\rangle=\left(  1\text{, }0\right)
^{\mathrm{T}}$ and $|B\rangle=\left(  a\text{, }b\right)  ^{\mathrm{T}}$. We
have,\noindent\
\begin{align}
U_{\mathrm{opt}}^{\mathrm{B}}(\tau_{\mathrm{min}})%
\begin{pmatrix}
1\\
0
\end{pmatrix}
&  =e^{i\text{\textrm{arg}}(a)}\left\{  \sqrt{1-|b|^{2}}%
\begin{pmatrix}
1\\
0
\end{pmatrix}
-i|b|\left[  \cos({\theta)}%
\begin{pmatrix}
0\\
1
\end{pmatrix}
+\sin({\theta)}%
\begin{pmatrix}
0\\
i
\end{pmatrix}
\right]  \right\} \nonumber\label{eq83}\\
&  =e^{i\text{\textrm{arg}}(a)}%
\begin{pmatrix}
\sqrt{1-|b|^{2}}\\
-i|b|e^{i\theta}%
\end{pmatrix}
\nonumber\\
&  =e^{i\text{\textrm{arg}}(a)}%
\begin{pmatrix}
\sqrt{1-|b|^{2}}\\
-i|b|e^{i\left[  \text{\textrm{arg}}(b)-\text{\textrm{arg}}(a)-\frac{3}{2}%
\pi\right]  }%
\end{pmatrix}
\nonumber\\
&  =%
\begin{pmatrix}
|a|e^{i\text{\textrm{arg}}(a)}\\
|b|e^{i\text{\textrm{arg}}(b)}%
\end{pmatrix}
\nonumber\\
&  =%
\begin{pmatrix}
a\\
b
\end{pmatrix}
\text{,}%
\end{align}
which is exactly the column representation of the final state $|B\rangle$.
Therefore, this ends our explicit verification. Having presented a critical
revisitation of both Mostafazadeh's (see Eqs. (\ref{eq16}), (\ref{eq19}),
(\ref{eq30}), and (\ref{eq31})) and Bender's (see Eqs. (\ref{eq77}),
(\ref{carlo25}), and (\ref{carlo20})) approaches, we are now ready to bring to
light the analogies and peculiarities of these two approaches to optimal-speed
quantum evolutions on the Bloch sphere. For completeness, we summarize in
Table I the peculiarities of Mostafazadeh's and Bender's schemes for
optimal-speed quantum evolutions (with stationary Hamiltonians) on the Bloch
sphere.\begin{table}[t]
\centering
\begin{tabular}
[c]{c|c|c|c|c}\hline\hline
\textbf{Quantum} \textbf{construction} & \textbf{Initial state} &
\textbf{Final state} & \textbf{Hamiltonian} & \textbf{Optimization}\\\hline
Mostafazadeh & Arbitrary & Arbitrary & Traceless & Energy uncertainty\\\hline
Bender & North pole & Arbitrary & Not traceless & Evolution time\\\hline
\end{tabular}
\caption{Schematic description of the main features of Mostafazadeh's and
Bender's approaches to optimal-speed quantum evolutions on the Bloch sphere.}%
\end{table}

\section{Link between the two approaches}

In this section, we discuss the connection between Mostafazadeh's and Bender's
approaches. For completeness, we recall that the main results in
Mostafazadeh's approach appear in Eqs. (\ref{eq16}), (\ref{eq19}),
(\ref{eq30}), and (\ref{eq31}). For the main findings in the context of
Bender's approach, we make reference to Eqs. (\ref{eq77}), (\ref{carlo25}),
and (\ref{carlo20}).

Recalling that Mostafazadeh's approach is specified by arbitrary initial and
final states, we should extend Bender's analysis to the case $|A\rangle
=\left(  a_{1}\text{, }a_{2}\right)  ^{\mathrm{T}}$ with $|a_{1}|^{2}%
+|a_{2}|^{2}=1$ (since it was originally developed for an initial state at the
north pole of the Bloch sphere). However, extending the analysis carried out
for $|A\rangle=\left(  1\text{, }0\right)  ^{\mathrm{T}}$ to the case
$|A\rangle=\left(  a_{1}\text{, }a_{2}\right)  ^{\mathrm{T}}$ leads to very
complicated calculations. For this reason, we prefer to show that when we
extend Mostafazadeh's approach to non-traceless Hamiltonians (since it was
originally developed for traceless Hamiltonians) we exactly recover Bender's
result (which, instead, did not assume tracelessness) once we also assume (as
in Bender's approach) that $|A\rangle=\left(  1\text{, }0\right)
^{\mathrm{T}}$. This would show in an explicit manner the equivalence between
the two procedures, letting us conclude that the generalization of the
Mostafazadeh approach to non-traceless Hamiltonians exactly represents a
generalization (since it works for arbitrary initial and final states) of
Bender's approach.

Let us recall that a generic bidimensional Hamiltonian $\mathrm{H}$ can always
be decomposed as
\begin{equation}
\mathrm{H}=\mathrm{H}^{\prime}+\frac{\text{\textrm{Tr}}\left(
\text{\textrm{H}}\right)  }{2}\mathbf{1}\text{.}\label{eq84}%
\end{equation}
Since the Hamiltonian is $2\times2$, the eigenvalues will be of the form $a\pm
b$ with $b>0$. Since the trace can be calculated in any basis, we can
calculate it in the basis of the eigenvectors of $\mathrm{H}$ and obtain
\textrm{Tr}$\left(  \text{\textrm{H}}\right)  =2a=E_{+}+E_{-}$. Then, always
in the basis of the eigenvectors of $\mathrm{H}$, we get that $\mathrm{H}%
^{\prime}$ must be%
\begin{equation}
\mathrm{H}^{\prime}=%
\begin{pmatrix}
b & 0\\
0 & -b
\end{pmatrix}
\text{,}\label{eq85}%
\end{equation}
with $b=\left(  E_{+}-E_{-}\right)  /2$. At the same time, $a$ can be related
to the eigenvalues of the Hamiltonian by $a=(E_{+}+E_{-})/2$. Given that
\textrm{Tr}$\left(  \text{\textrm{H}}\right)  =2a=E_{+}+E_{-}$ and given the
decomposition in Eq. (\ref{eq84}), we get that the generic non-traceless
Hamiltonian with energy eigenvalues $E_{+}$ and $E_{-}$ is given by
\begin{equation}
\mathrm{H}=\frac{E_{+}+E_{-}}{2}\mathbf{1}+\frac{E_{+}-E_{-}}{2}\vec{n}%
\cdot\vec{\sigma}.\label{eq86}%
\end{equation}
We note that in Mostafazadeh's analysis, the term $\vec{n}\cdot\vec{\sigma}$
in Eq. (\ref{eq86}) corresponds to
\begin{equation}
i\cot{\frac{\theta}{2}}\left(  \frac{|\psi_{F}\rangle\langle\psi_{I}|}%
{\langle\psi_{I}|\psi_{F}\rangle}-\frac{|\psi_{I}\rangle\langle\psi_{F}%
|}{\langle\psi_{F}|\psi_{I}\rangle}\right)  \text{,}\label{eq87}%
\end{equation}
with $\theta$ being the angular distance on the Bloch sphere between the
states $|\psi_{I}\rangle$ and $|\psi_{F}\rangle$. Therefore, the
generalization of the Hamiltonian $\mathrm{H}_{\mathrm{opt}}^{\mathrm{M}}$ in
Eq. (\ref{eq16}) with generic eigenvalues $E_{+}$ and $E_{-}$ becomes
\begin{equation}
\mathrm{H}_{\mathrm{opt}}^{\mathrm{M}}=\frac{E_{+}+E_{2}}{2}\mathbf{1}%
+i\frac{E_{+}-E_{2}}{2}\cot({\frac{\theta}{2})}\left(  \frac{|\psi_{F}%
\rangle\langle\psi_{I}|}{\langle\psi_{I}|\psi_{F}\rangle}-\frac{|\psi
_{I}\rangle\langle\psi_{F}|}{\langle\psi_{F}|\psi_{I}\rangle}\right)
\text{.}\label{eq88}%
\end{equation}
According to Eq. (\ref{carlo13}), the maximum energy uncertainty (which,
relying on Mostafazadeh's result, corresponds to the optimal time evolution)
is obtained when $|c_{+}|^{2}=1/2$. In the context of Bender's analysis, this
constraint becomes%
\begin{equation}
\Delta E_{\psi}=\frac{(E_{+}-E_{-})^{2}}{4}=\frac{\omega^{2}}{4}%
\text{,}\label{eq89}%
\end{equation}
where $\omega\overset{\text{def}}{=}E_{+}-E_{-}=\sqrt{(s-u)^{2}+4r^{2}}$ as in
Eq. (\ref{eq67}). We can see from Eq. (\ref{eq89}) how, when the purpose is
the search of the optimal time Hamiltonian, imposing a constraint on the
difference between the eigenvalues of the Hamiltonian is analogous to imposing
a constraint on the standard deviation of $\mathrm{H}$. This is already an
important clue of the equivalence between the two descriptions (i.e.,
Mostafazadeh's and Bender's approaches). However, we can make this equivalence
even more explicit.\newline Let us consider Bender's assumptions for initial,
final states, and energy dispersion. We have,
\begin{equation}
|\psi_{I}\rangle=|0\rangle\text{, }|\psi_{F}\rangle=a|0\rangle+b|1\rangle
\text{, }E_{+}-E_{-}=\omega\text{, and }E_{+}+E_{-}=s+u\text{.}\label{carlo22}%
\end{equation}
Representing $|\psi_{I}\rangle$ and $|\psi_{F}\rangle$ on the Bloch sphere, we
notice that $|\psi_{I}\rangle$ corresponds to the $z$-versor. Then, the angle
$\theta$ between $|\psi_{I}\rangle$ and $|\psi_{F}\rangle$, adopting the usual
representation of Bloch vectors $|\psi\rangle=\cos({\frac{\theta}{2}%
)}|0\rangle+\sin({\frac{\theta}{2})}e^{i\phi}|1\rangle$, is related to the
coefficients $a$ and $b$ by%
\begin{equation}
|a|=\cos({\frac{\theta}{2})}\text{, and }|b|=\sin({\frac{\theta}{2})}%
\text{,}\label{eq91}%
\end{equation}
respectively. For clarity, we remark once again that the angle $\theta$ in Eq.
(\ref{eq91}) differs from the parameter $\theta$ in Eqs. (\ref{eq64}) and
(\ref{carlo18}) used in Bender's approach as we clarified in the previous
section. From Eq. (\ref{eq91}), we get
\begin{equation}
\cot({\frac{\theta}{2})}=\frac{|a|}{|b|}\text{.}\label{eq92}%
\end{equation}
At the same time, from Eq. (\ref{carlo22}), we also have
\begin{equation}
\left\langle \psi_{I}\left\vert \psi_{F}\right.  \right\rangle =a\text{,
}\left\langle \psi_{F}\left\vert \psi_{I}\right.  \right\rangle =a^{\ast
}\text{, }|\psi_{F}\rangle\langle\psi_{I}|=a|0\rangle\langle0|+b|1\rangle
\langle0|\text{, and }|\psi_{I}\rangle\langle\psi_{F}|=a^{\ast}|0\rangle
\langle0|+b^{\ast}|0\rangle\langle1|\text{.}\label{carlo23}%
\end{equation}
Assuming Bender's optimal result, we also have that $E_{+}+E_{-}=s+u$.
Inserting Eqs. (\ref{carlo22}), (\ref{eq92}), and (\ref{carlo23}) into Eq.
(\ref{eq88}), the Mostafazadeh generalized expression for the optimal
Hamiltonian becomes
\begin{align}
\mathrm{H}_{\mathrm{opt}}^{\mathrm{M}} &  =\frac{s+u}{2}\mathbf{1}%
+i\frac{\omega}{2}\frac{|a|}{|b|}\left(  \frac{a|0\rangle\langle
0|+b|1\rangle\langle0|}{a}-\frac{a^{\ast}|0\rangle\langle0|+b^{\ast}%
|0\rangle\langle1|}{a^{\ast}}\right)  \nonumber\\
&  =\frac{s+u}{2}\mathbf{1}+i\frac{\omega}{2}\frac{|a|}{|b|}\left(  \frac
{b}{a}|1\rangle\langle0|-\frac{b^{\ast}}{a^{\ast}}|0\rangle\langle1|\right)
\nonumber\\
&  =\frac{s+u}{2}\mathbf{1}+i\frac{\omega}{2}\left[  e^{i\left[
\text{\textrm{arg}}(b)-\text{\textrm{arg}}(a)\right]  }|1\rangle
\langle0|-e^{-i\left[  \text{\textrm{arg}}(b)-\text{\textrm{arg}}(a)\right]
}|0\rangle\langle1|\right]  \text{,}\label{carlo24}%
\end{align}
where in the last line of Eq. (\ref{carlo24}) we used the identity$\left(
|a|b\right)  /\left(  a|b|\right)  $ $=e^{i(\left[  \text{arg}(b)-\text{arg}%
(a)\right]  }$. In the end, the matrix representation of $\mathrm{H}%
_{\mathrm{opt}}^{\mathrm{M}}$ in Eq. (\ref{carlo24}) with respect to the
canonical computational basis becomes
\begin{equation}
\mathrm{H}_{\mathrm{opt}}^{\mathrm{M}}=%
\begin{pmatrix}
\frac{s+u}{2} & \frac{\omega}{2}e^{-i\left[  \text{arg}(b)-\text{arg}%
(a)-\frac{3}{2}\pi\right]  }\\
\frac{\omega}{2}e^{i\left[  \text{arg}(b)-\text{arg}(a)-\frac{3}{2}\pi\right]
} & \frac{s+u}{2}%
\end{pmatrix}
=\mathrm{H}_{\mathrm{opt}}^{\mathrm{B}}\text{,}\label{eq95}%
\end{equation}
which is exactly the optimal Hamiltonian $\mathrm{H}_{\mathrm{opt}%
}^{\mathrm{B}}$ in Eq. (\ref{eq77}) once we set\textbf{ }$s=u=-(\omega
/2)\left[  \mathrm{arg}(a)/\arcsin{|b|}\right]  $. Finally, the minimum time
found by Mostafazadeh is $\tau_{\mathrm{\min}}^{\mathrm{M}}=(\hbar s)/E$ with
$s$ denoting the geodesic distance according to the Fubini-Study metric.
Therefore, $\tau_{\mathrm{\min}}^{\mathrm{M}}$ can be recast as
\begin{equation}
\tau_{\mathrm{\min}}^{\mathrm{M}}=\frac{\hbar\arccos{(|\langle\psi_{I}%
|\psi_{F}\rangle|}}{E}=\frac{2\hbar\arccos{|\langle\psi_{I}|\psi_{F}\rangle|}%
}{E_{+}-E_{-}}\text{.}\label{eq96}%
\end{equation}
If we plug Bender's states with $\langle\psi_{I}|\psi_{F}\rangle=a$ inside the
expression for $\tau_{\mathrm{\min}}^{\mathrm{M}}$ in Eq. (\ref{eq96}), we
find
\begin{equation}
\tau_{\mathrm{\min}}^{\mathrm{M}}=\frac{2\hbar\arccos\sqrt{1-|b|^{2}}}{\omega
}=\frac{2\hbar}{\omega}\arcsin{|b|=}\tau_{\mathrm{\min}}^{\mathrm{B}}%
\text{.}\label{eq97}%
\end{equation}
Eq. (\ref{eq97}) shows that $\tau_{\mathrm{\min}}^{\mathrm{M}}=\tau
_{\mathrm{\min}}^{\mathrm{B}}$ (with $\tau_{\mathrm{\min}}^{\mathrm{M}}$ and
$\tau_{\mathrm{\min}}^{\mathrm{B}}$ in Eqs. (\ref{eq5}) and (\ref{eq72}),
respectively) and ends our comparative analysis of the two approaches to
optimal-speed quantum evolutions. At a foundational level, the link between
maximizing the energy uncertainty in Mostafazadeh's approach and minimizing
the evolution time in Bender's approach is rooted in suitable time-energy
inequality constraints that govern a quantum evolution between orthogonal
and/or non-orthogonal quantum states \cite{anandan90,cafaro21QR}. In summary,
we reiterate that Mostafazadeh's original approach in Ref. \cite{ali09} takes
into consideration generic initial and final states but assumes a traceless
Hamiltonian. Bender's original approach in Ref. \cite{bender07}, instead, does
not assume a traceless Hamiltonian. However, it does consider a particular
initial state (i.e., the north pole on the Bloch sphere). What we showed in
Eq. (\ref{eq95}) is that Mostafazadeh's and Bender's approaches are equivalent
when we extend Mostafazadeh's approach to Hamiltonians with nonzero trace and,
at the same time, focus on an initial quantum state on the north pole of the
Bloch sphere. We are now ready for our summary and final remarks.

\section{Conclusions}

In this paper, we presented a comparative analysis between two alternative
constructions of optimal-speed Hamiltonian quantum evolutions for two-level
quantum systems. In the first construction (i.e., Mostafazadeh's approach
\cite{ali09}), one seeks the optimal-speed Hamiltonian between arbitrary
initial and final quantum states. This construction assumes a traceless
Hamiltonian and the quantity to be maximized during the evolution is the
energy uncertainty (i.e., the standard deviation of the Hamiltonian operator
calculated with respect to the initial state of the system). Expressions of
the optimal Hamiltonian $\mathrm{H}_{\mathrm{opt}}^{\mathrm{M}}$ and its
corresponding unitary time evolution operator $U_{\mathrm{opt}}^{\mathrm{M}%
}(t)$ appear in Eqs. (\ref{eq16}) and (\ref{eq19}), respectively.
Interestingly, we also found (see Eqs. (\ref{eq30}) and (\ref{eq31})) an
alternative geometrically meaningful expression of these operators in terms of
the unit Bloch vectors $\hat{a}$ and $\hat{b}$ corresponding to the initial
and final states, respectively. More specifically, we showed in an explicit
manner that the time optimal evolution on the Bloch sphere given by the
Hamiltonian $\mathrm{H}_{\mathrm{opt}}^{\mathrm{M}}$ in Eq. (\ref{eq16})
results in a rotation around the axis $(\hat{a}\times\hat{b})/\sqrt{(\hat
{a}\times\hat{b})\cdot(\hat{a}\times\hat{b})}$. In the second construction
(i.e., Bender's approach \cite{bender07}), one seeks the optimal-speed
Hamiltonian between an initial state located at the north pole on the Bloch
sphere and an arbitrary final quantum state. This construction does not assume
a traceless Hamiltonian and the quantity to be minimized during the quantum
motion is the evolution time subject to the constraint that the difference
between the largest ($E_{+}$) and smallest ($E_{-}$) eigenvalues of the
Hamiltonian is kept fixed. Expressions of the optimal Hamiltonian
$\mathrm{H}_{\mathrm{opt}}^{\mathrm{B}}$ and its corresponding unitary time
evolution operator $U_{\mathrm{opt}}^{\mathrm{B}}(t)$ appear in Eqs.
(\ref{eq77}) and (\ref{carlo20}), respectively. After discussing the two
approaches separately, we discussed their connection. Specifically, we pointed
out that the generalization of any one of the two procedures would be very
tedious if performed from the start in the construction. In the Mostafazadeh's
approach, the generalization requires not assuming a traceless Hamiltonian. In
Bender's approach, the generalization requires not assuming an initial state
located at the north pole of the Bloch sphere. However, we were able to show
(see Eq. (\ref{eq95})) that Mostafazadeh's and Bender's approaches are
equivalent when we extend Mostafazadeh's approach to Hamiltonians with nonzero
trace and, at the same time, focus on an initial quantum state placed on the
north pole of the Bloch sphere. In both approaches, the key geometric feature
is that the optimal unitary evolution operator is essentially a rotation about
an axis that is orthogonal to the unit Bloch vectors that correspond to the
initial and final quantum states of the system. Stated otherwise, for
optimal-time stationary Hamiltonian evolutions of qubits, the optimal magnetic
field configuration for evolving a quantum system along a geodesic path in
projective Hilbert space is specified by a magnetic field that is orthogonal
to both the initial and final unit Bloch vectors of the system.\bigskip

In addition to its clear pedagogical nature, we think our work is relevant for
at least two innovative lines of research. First, it serves as a natural
background for geometrically quantifying deviations from geodesic quantum
motion on the Bloch sphere. These deviations, in turn, are expected to
generate quantum evolutions with nonzero curvature worth of additional
attention \cite{brody96,brody13,laba17,laba22,alsing24A,alsing24B,luba24}.
Second, it would help extending the study on the complexity of quantum
evolutions for systems violating the geodesicity property
\cite{cafaroprd22,cafaropre22} and, possibly, going beyond single-qubit pure
states. In an exploratory step, for instance, one might be interested in
observing what happens in the Bloch sphere with mixed states
\cite{sam95A,sam95B,carloepj,carloPRA,carloQR}, for instance. Despite its
relative simplicity, studying the complexity of nongeodesic paths on the Bloch
sphere, of geodesic paths on deformed Bloch spheres, and of nongeodesic paths
in the Bloch sphere can be rather challenging
\cite{brown19,chapman18,ruan21,huang24}. Ultimately, one might be concerned
with characterizing in an explicit way deviations from the geodesicity
property of quantum evolutions beyond two-level quantum systems, in both
unitary and nonunitary settings. This would be an achievement of great value
since, to the best of our knowledge, there are only algorithms capable of
solving unconstrained quantum brachistochrone problems in a unitary setting
focused on minimally energy wasteful paths
\cite{uzdin12,russell17,uzdin15,suri18,campa19,xu24} that connect two
isospectral mixed quantum states in terms of fast evolutions
\cite{campaioli19,dou23,carlini08,campbell21,hornedal22,nade24}.\bigskip

For the time being, we leave a more in-depth quantitative discussion on these
potential geometric extensions of our analytical findings, including
generalizations to mixed state geometry and nongeodesic quantum evolutions in
higher-dimensional Hilbert spaces, to forthcoming scientific investigations.

\begin{acknowledgments}
The authors thank two anonymous reviewers for useful comments leading to an
improved version of this manuscript. Any opinions, findings and conclusions or
recommendations expressed in this material are those of the author(s) and do
not necessarily reflect the views of their home Institutions.
\end{acknowledgments}

\pagebreak

\appendix

\section{Traceless optimal Hamiltonians}

In this Appendix, we justify the reason behind the choice of working solely
with traceless Hamiltonians and we show how working with non-traceless
Hamiltonians would have generated identical results in Mostafazadeh's approach.

Suppose to have a general Hamiltonian $\mathrm{H}$ for a two-level quantum
system. According to the spectral decomposition theorem, the Hamiltonian can
always be written as
\begin{equation}
\mathrm{H}=E_{+}|E_{+}\rangle\langle E_{+}|+E_{-}|E_{-}\rangle\langle
E_{-}|\text{,} \label{eq32}%
\end{equation}
with $E_{+}$ and $E_{-}$ corresponding to its highest and lowest eigenvalues,
respectively. Suppose that one wishes to calculate the standard deviation
$\Delta E_{\psi}$ of the Hamiltonian $\mathrm{H}$ on a generic normalized
state $|\psi\rangle$,
\begin{equation}
\Delta E_{\psi}\overset{\text{def}}{=}\sqrt{\langle\psi|\mathrm{H}^{2}%
|\psi\rangle-\langle\psi|\mathrm{H}|\psi\rangle^{2}}\text{.} \label{eq33}%
\end{equation}
Since the eigenstates of $\mathrm{H}$ constitute an orthonormal basis, we can
decompose $|\psi\rangle$ as
\begin{equation}
|\psi\rangle=c_{+}|E_{+}\rangle+c_{-}|E_{-}\rangle\text{,} \label{eq34}%
\end{equation}
with $|c_{+}|^{2}+|c_{-}|^{2}=1$. Moreover, thanks to the spectral
decomposition in Eq. (\ref{eq32}), we get
\begin{equation}
\mathrm{H}^{2}=E_{+}^{2}|E_{+}\rangle\langle E_{+}|+E_{-}^{2}|E_{-}%
\rangle\langle E_{-}|\text{.} \label{eq35}%
\end{equation}
Then, plugging Eqs. (\ref{eq32}), (\ref{eq34}) and (\ref{eq35}) into Eq.
(\ref{eq33}), after some algebra we obtain%
\begin{equation}
\Delta E_{\psi}=(E_{+}-E_{-})\sqrt{|c_{+}|^{2}-|c_{+}|^{4}}\text{.}
\label{eq36b}%
\end{equation}
As a side remark, we note that $\Delta E_{\psi}$ in Eq. (\ref{eq36b}) can also
be expressed in terms of $|c_{-}|$ because of the normalization condition
$|c_{+}|^{2}+|c_{-}|^{2}=1$. Then, we can see from Eq. (\ref{eq36b}) that,
apart from the particular state $|\psi\rangle$ on which we choose to calculate
$\Delta E_{\psi}$, the value of the standard deviation only depends on the
difference of the eigenstates of the Hamiltonian. Moreover, if we consider a
traceless Hamiltonian, we exactly recover the result originally found by
Mostafazadeh and represented here in Eq. (\ref{eq4}). Indeed, starting from
Eq. (\ref{eq4}), we get
\begin{align}
\Delta E_{\psi}  &  =E\sqrt{1-{\left(  \frac{|c_{1}|^{2}-|c_{2}|^{2}}%
{|c_{1}|^{2}+|c_{2}|^{2}}\right)  }^{2}}\nonumber\\
&  =E\sqrt{1-{\frac{|c_{1}|^{4}+|c_{2}|^{4}-2|c_{1}|^{2}|c_{2}|^{2}}%
{|c_{1}|^{4}+|c_{2}|^{4}+2|c_{1}|^{2}|c_{2}|^{2}}}}\nonumber\\
&  =2E\sqrt{|c_{1}|^{2}-|c_{1}|^{4}}\text{.} \label{carlo13}%
\end{align}
Eq. (\ref{carlo13}) represents exactly the same result as in Eq.
(\ref{eq36b}), since for traceless Hamiltonians the eigenvalues must be $E$
and $-E$ and, therefore, $E_{+}-E_{-}=2E$. Notice now that an arbitrary
Hamiltonian $\mathrm{H}$ of a two-level quantum system can always be
decomposed as
\begin{equation}
\mathrm{H}=\mathrm{H}^{\prime}+\frac{\text{\textrm{Tr}}\left(  \mathrm{H}%
\right)  }{2}\mathbf{1}\text{,} \label{eq38}%
\end{equation}
with $\mathrm{H}^{\prime}$ being traceless. In particular, the dynamics
generated by the two Hamiltonians $\mathrm{H}$ and $\mathrm{H}^{\prime}$ in
the projective Hilbert space are the same since
\begin{equation}
e^{-\frac{i}{\hbar}\mathrm{H}t}=e^{-\frac{i}{\hbar}\mathrm{H}^{\prime}%
t}e^{-\frac{i}{\hbar}\frac{\text{\textrm{Tr}}\left(  \mathrm{H}\right)  }%
{2}\mathbf{1}t}\text{,} \label{eq39}%
\end{equation}
where $e^{-\frac{i}{\hbar}\frac{\text{\textrm{Tr}}\left(  \mathrm{H}\right)
}{2}\mathbf{1}t}$ results in a simple (global) phase factor. Moreover, when we
consider bidimensional Hamiltonians, we have that the eigenvalues are
solutions of a second degree polynomial equation and must have the form $a\pm
b$ with $b\geq0$. Since the trace of a matrix can be calculated in any basis,
we can calculate it in the basis of eigenvectors of $\mathrm{H}$ and obtain
\textrm{Tr}$\left(  \mathrm{H}\right)  =2a=E_{+}+E_{-}$. Moreover,
$\mathrm{H}^{\prime}$ can be explicitly recast as
\begin{equation}
\mathrm{H}^{\prime}=%
\begin{pmatrix}
b & 0\\
0 & -b
\end{pmatrix}
\text{,} \label{eq40}%
\end{equation}
with $b=(E_{+}-E_{-})/2$. We can clearly see from Eq. (\ref{eq40}) that the
traceless Hamiltonian $\mathrm{H}^{\prime}$ only depends on the difference of
the eigenvalues of the original Hamiltonian \textrm{H}. Then, since the
standard deviation of a generic Hamiltonian only depends on such difference as
well as shown in Eq. (\ref{eq36b}), if the purpose is to maximize the standard
deviation (exactly like in Mostafazadeh's approach) then we can focus on the
traceless Hamiltonian $\mathrm{H}^{\prime}$ instead of \textrm{H. }Moreover,
because of Eq. (\ref{eq39}), we also know that the quantum evolution in
projective Hilbert is the same, regardless of which Hamiltonian we use (be it
\textrm{H} or \textrm{H}$^{\prime}$). In conclusion, the choice of focusing on
traceless Hamiltonians does not affect the generality of the result on finding
optimal-speed quantum Hamiltonian evolutions and is fully justified.


\begin{thebibliography}{99}                                                                                               %


\bibitem {karol}I. Bengtsson and K. Zyczkowski, \emph{Geometry of Quantum
States}, Cambridge University Press (2006).

\bibitem {bohm03}A. Bohm, A. Mostafazadeh, H. Koizumi, Q. Niu, and J.
Zwanziger, \emph{The Geometric Phase in Quantum Systems}, Springer Berlin,
Heidelberg (2003).

\bibitem {dario04}D. Chruscinski and A. Jamiolkowski, \emph{Geometric Phases
in Classical and Quantum Mechanics}, Birkh\"{a}user Boston (2004).

\bibitem {avron09}J. E. Avron and O. Kenneth, \emph{Entanglement and the
geometry of two qubits}, Ann. Phys. \textbf{324}, 470 (2009).

\bibitem {avron20}J. Avron and O. Kenneth, \emph{An elementary introduction to
the geometry of quantum states with pictures}, Rev. Math. Phys. \textbf{32},
2030001 (2020).

\bibitem {anza21}F. Anza and J. P. Crutchfield, \emph{Beyond density matrices:
Geometric quantum states}, Phys. Rev. \textbf{A103}, 062218 (2021).

\bibitem {anza22}F. Anza and J. P. Crutchfield, \emph{Quantum information
dimension and geometric entropy}, PRX Quantum \textbf{3}, 020355 (2022).

\bibitem {anza22B}F. Anza and J. P. Crutchfield, \emph{Geometric quantum
thermodynamics}, Phys. Rev. \textbf{E106}, 054102 (2022).

\bibitem {anza24}F. Anza and J. P. Crutchfield, \emph{Maximum geometric
quantum entropy}, Entropy \textbf{26}, 225 (2024).

\bibitem {beggs20}E. J. Beggs and S. Majid, \emph{Quantum Riemannian
Geometry}, Springer Nature Switzerland AG (2020).

\bibitem {majid19}S. Majid, \emph{Quantum gravity on a square graph}, Class.
Quantum Grav. \textbf{36}, 245009 (2019).

\bibitem {beggs21}E. J. Beggs and S. Majid, \emph{Quantum geodesics in quantum
mechanics}, arXiv:math-ph/1912.13376 (2021).

\bibitem {beggs22}E. J. Beggs and S. Majid, \emph{Quantum geodesic flows and
curvature}, Lett. Math. Phys. \textbf{113}, 73 (2023).

\bibitem {wootters81}W. K. Wootters, \emph{Statistical distance and Hilbert
space}, Phys. Rev. \textbf{D23}, 357 (1981).

\bibitem {braunstein94}S. L. Braunstein and C. M. Caves, \emph{Statistical
distance and the geometry of quantum states}, Phys. Rev. Lett. \textbf{72},
3439 (1994).

\bibitem {page87}D. N. Page, \emph{Geometrical description of Berry's phase},
Phys. Rev. \textbf{A36}, 3479(R) (1987).

\bibitem {abe92}S. Abe,\emph{\ Quantum-state metric and correlations}, Phys.
Rev. \textbf{A46}, 1667 (1992).

\bibitem {boscain06}U. Boscain and P. Mason, \emph{Time minimal trajectories
for a spin }$1/2$\emph{\ particle in a magnetic field}, J. Math. Phys.
\textbf{47}, 062101 (2006).

\bibitem {boozer12}A. D. Boozer, \emph{Time-optimal synthesis of }%
$SU(2)$\emph{\ transformations for a spin-}$1/2$\emph{\ system}, Phys. Rev.
\textbf{A85}, 012317 (2012).

\bibitem {fry08}A. M. Frydryszak and V. M. Tkachuk, \emph{Quantum
brachistochrone problem for a spin-}$1$\emph{\ system in a magnetic field},
Phys. Rev. \textbf{A77}, 014103 (2008).

\bibitem {kuz15}A. R. Kuzmak and V. M. Tkachuk, \emph{The quantum
brachistochrone problem for an arbitrary spin in a magnetic field}, Phys.
Lett. \textbf{A379}, 1233 (2015).

\bibitem {braunstein95}S. L. Braunstein and C. M. Caves, \emph{Geometry of
quantum states}. In: B. V. Belavkin, O. Hirota, and R. L. Hudson (eds.),
Quantum Communications and Measurement, pp. 21-30. Springer, Boston, MA (1995).

\bibitem {brody03}D. C. Brody, \emph{Elementary derivation for passage times},
J. Phys. A: Math. Gen. \textbf{36}, 5587 (2003).

\bibitem {carlini06}A. Carlini, A. Hosoya, T. Koike, and Y. Okudaira,
\emph{Time-optimal quantum evolution}, Phys. Rev. Lett. \textbf{96}, 060503 (2006).

\bibitem {brody06}D. C. Brody and D. W. Hook, \emph{On optimum Hamiltonians
for state transformations}, J. Phys. A: Math. Gen. \textbf{39}, L167 (2006).

\bibitem {brody07}D. C. Brody and D. W. Hook, \emph{On optimum Hamiltonians
for state transformation}, J. Phys. A: Math. Theor. \textbf{40}, 10949 (2007).

\bibitem {bender07}C. M. Bender, D. C. Brody, H. F. Jones, and B. K. Meister,
\emph{Faster than Hermitian quantum mechanics}, Phys. Rev. Lett. \textbf{98},
040403 (2007).

\bibitem {bender09}C. M. Bender and D. C. Brody, \emph{Optimal time evolution
for Hermitian and non-Hermitian Hamiltonians}, Lecture Notes in Physics
\textbf{789}, 341 (2009).

\bibitem {uhlmann92}A. Uhlmann,\ \emph{An energy dispersion estimate}, Phys.
Lett. \textbf{A161}, 329 (1992).

\bibitem {ali09}A. Mostafazadeh, \emph{Hamiltonians generating optimal-speed
evolutions}, Phys. Rev. \textbf{A79}, 014101 (2009).

\bibitem {cafaro23}C. Cafaro and P. M. Alsing, \emph{Qubit geodesics on the
Bloch sphere from optimal-speed Hamiltonian evolutions}, Class. Quantum Grav.
\textbf{40}, 115005 (2023).

\bibitem {anandan90}J. Anandan and Y. Aharonov, \emph{Geometry of quantum
evolution}, Phys. Rev. Lett. \textbf{65}, 1697 (1990).

\bibitem {cafaro20}C. Cafaro, S. Ray, and P. M. Alsing, \emph{Geometric
aspects of analog quantum search evolutions}, Phys. Rev. \textbf{A102, }052607 (2020).

\bibitem {hamilton23}G. A. Hamilton and B. K. Clark, \emph{Quantifying unitary
flow efficiency and entanglement for many-body localization}, Phys. Rev.
\textbf{B107}, 064203 (2023).

\bibitem {wolf59}E. Wolf, \emph{Coherence properties of partially polarized
electromagnetic radiation}, Il Nuovo Cimento \textbf{13}, 1180 (1959).

\bibitem {wolf07}E. Wolf,\emph{ Introduction to the Theory of Coherence and
Polarization of Light}, Cambridge University Press (2007).

\bibitem {cafaro22}C. Cafaro\textbf{,} S. Ray, and P. M. Alsing,
\emph{Optimal-speed unitary quantum time evolutions and propagation of light
with maximal degree of coherence}, Phys. Rev. \textbf{A105}, 052425 (2022).

\bibitem {provost80}J. P. Provost and G. Vallee, \emph{Riemannian structure on
manifolds of quantum states}, Commun. Math. Phys. \textbf{76}, 289 (1980).

\bibitem {mukunda93}N. Mukunda and R. Simon, \emph{Quantum kinematic approach
to the geometric phase I. General Formalism}, Annals of Physics \textbf{228},
205 (1993).

\bibitem {crell09}A. Uhlmann and B. Crell, \emph{Geometry of state spaces},
Lecture Notes in Physics \textbf{768}, 1 (2009).

\bibitem {cafaro21QR}C. Cafaro and P. M.\ Alsing, \emph{Minimum time for the
evolution to a nonorthogonal quantum state and upper bound of the geometric
efficiency of quantum evolutions}, Quantum Reports \textbf{3}, 444 (2021).

\bibitem {brody96}D. B. Brody and L. P. Hughston, \emph{Geometry of quantum
statistical inference}, Phys. Rev. Lett. \textbf{77}, 2851 (1996).

\bibitem {brody13}D. C. Brody and Eva-Marie Graefe, \emph{Information geometry
of complex Hamiltonians and exceptional points}, Entropy \textbf{15}, 3361 (2013).

\bibitem {laba17}H. P. Laba and V. M. Tkachuk, \emph{Geometric characteristics
of quantum evolution: Curvature and torsion}, Condensed Matter Physics
\textbf{20}, 13003 (2017).

\bibitem {laba22}Kh. P. Gnatenko, H. P. Laba, and V. M. Tkachuk,
\emph{Geometric properties of evolutionary graph states and their detection on
a quantum computer}, Phys. Lett. \textbf{A452}, 128434 (2022).

\bibitem {alsing24A}P. M.\ Alsing and C. Cafaro, \emph{From the classical
Frenet--Serret apparatus to the curvature and torsion of quantum-mechanical
evolutions. Part I. Stationary Hamiltonians}, Int. J. Geom. Methods Mod. Phys.
\textbf{21}, 2450152 (2024).

\bibitem {alsing24B}P. M.\ Alsing and C. Cafaro, \emph{From the classical
Frenet--Serret apparatus to the curvature and torsion of quantum-mechanical
evolutions. Part II. Nonstationary Hamiltonians}, Int. J. Geom. Methods Mod.
Phys. \textbf{21}, 2450151 (2024).

\bibitem {luba24}C. Mc Keever and M. Lubasch, \emph{Towards adiabatic quantum
computing using compressed quantum circuits}, PRX Quantum \textbf{5}, 020362 (2024).

\bibitem {cafaroprd22}C. Cafaro and P. M. Alsing, \emph{Complexity of pure and
mixed qubit geodesic paths on curved manifolds,} Phys. Rev. \textbf{D106,
}096004 (2022).

\bibitem {cafaropre22}C. Cafaro, S. Ray, and P. M. Alsing, \emph{Complexity
and efficiency of minimum entropy production probability paths from quantum
dynamical evolutions}, Phys. Rev. \textbf{E105}, 034143 (2022).

\bibitem {sam95A}S. L. Braunstein and C. M. Caves, \emph{Geometry of quantum
states}, Annals of the New York Academy of Sciences \textbf{755}, 786 (1995).

\bibitem {sam95B}S. L. Braunstein and C. M. Caves, \emph{Geometry of quantum
states}. In: B. V. Belavkin, O. Hirota, and R. L. Hudson (Eds.), Quantum
Communications and Measurement, pp. 21-30. Springer, Boston, MA (1995).

\bibitem {carloepj}C. Cafaro and P. M. Alsing, \emph{Bures and Sj\"{o}qvist
metrics over thermal state manifolds for spin qubits and superconducting flux
qubits}, Eur. Phys. J. Plus \textbf{138}, 655 (2023).

\bibitem {carloPRA}P. M. Alsing, C. Cafaro, O. Luongo, C. Lupo, S. Mancini,
and H. Quevedo, \emph{Comparing metrics for mixed quantum states: Sj\"{o}qvist
and Bures}, Phys. Rev. \textbf{A107}, 052411 (2023).

\bibitem {carloQR}P. M. Alsing, C. Cafaro, D. Felice, and O. Luongo,
\emph{Geometric aspects of mixed quantum states inside the Bloch sphere},
Quantum Reports \textbf{6}, 90 (2024).

\bibitem {brown19}A. R. Brown and L. Susskind, \emph{Complexity geometry of a
single qubit}, Phys. Rev. \textbf{D100}, 046020 (2019).

\bibitem {chapman18}S. Chapman, M. P. Heller, H. Marrachio, and F. Pastawski,
\emph{Toward a definition of complexity for quantum field theory states},
Phys. Rev. Lett. \textbf{120}, 121602 (2018).

\bibitem {ruan21}S.-M. Ruan, \emph{Circuit Complexity of Mixed States}, Ph.D.
in Physics, University of Waterloo (2021).

\bibitem {huang24}J.-H. Huang, S.-S. Dong, G.-L. Chen, N.-R. Zhou, F.-Y. Liu,
and L.-G. Qin,\ \emph{Shortest evolution path between two mixed states and its
realization}, Phys. Rev. \textbf{A109}, 042405 (2024).

\bibitem {uzdin12}R. Uzdin, U. G\"{u}nther, S. Rahav, and N. Moiseyev,
\emph{Time-dependent Hamiltonians with 100\% evolution speed efficiency}, J.
Phys. A: Math. Theor. \textbf{45}, 415304 (2012).

\bibitem {russell17}B. Russell and S. Stepney, \emph{The geometry of speed
limiting resources in physical models of computation}, Int. J. Found. Computer
Science \textbf{28}, 321 (2017).

\bibitem {uzdin15}R. Uzdin, A. Levy, and R. Kosloff, \emph{Equivalence of
quantum heat machines, and quantum-thermodynamic signatures}, Phys. Rev.
\textbf{X5}, 031044 (2015).

\bibitem {suri18}N. Suri, F. C. Binder, B. Muralidharan, and S. Vinjanampathy,
\emph{Speeding up thermalisation via open quantum system variational
optimisation}, Eur. Phys. J. Spec. Top. \textbf{227}, 203 (2018).

\bibitem {campa19}F. Campaioli, F. A. Pollock, and K. Modi, \emph{Tight,
robust, and feasible quantum speed limits for open dynamics}, Quantum
\textbf{3}, 168 (2019).

\bibitem {xu24}J. Xu et \emph{al}., \emph{Balancing the quantum speed limit
and instantaneous energy cost in adiabatic quantum evolution}, Chinese Phys.
Lett. \textbf{41}, 040202 (2024).

\bibitem {campaioli19}F. Campaioli, W. Sloan, K. Modi, and F. A. Pollok,
\emph{Algorithm for solving unconstrained unitary quantum brachistochrone
problems}, Phys. Rev. \textbf{A100}, 062328 (2019).

\bibitem {dou23}F.-Q. Dou, M.-P. Han, and C.-C. Shu, \emph{Quantum speed limit
under brachistochrone evolution}, Phys. Rev. Applied \textbf{20}, 014031 (2023).

\bibitem {carlini08}A. Carlini, A. Hosoya, T. Koike, and Y. Okudaira,
\emph{Time optimal quantum evolution of mixed states}, J. Phys. A: Math.
Theor. \textbf{41}, 045303 (2008).

\bibitem {campbell21}E. O'Connor, G. Guarnieri, and S. Campbell, \emph{Action
quantum speed limits}, Phys. Rev. \textbf{A103}, 022210 (2021).

\bibitem {hornedal22}N. Hornedal, D. Allan, and O. Sonnerborn,
\emph{Extensions of the Mandelstam-Tamm quantum speed limit to systems in
mixed states}, New J. Phys. \textbf{24}, 055004 (2022).

\bibitem {nade24}A. Naderzadeh-ostad and S. J. Akhtarshenas, \emph{Optimal
quantum speed for mixed states}, J. Phys. \textbf{A}: Math. Theor. 57, 075301 (2024).
\end{thebibliography}
\end{document}